\documentclass[10pt,a4paper]{article}

\usepackage{amsfonts,amsmath,amssymb,mathtools,mathrsfs,bbm,float}
\usepackage[ruled, vlined, linesnumbered]{algorithm2e}
\usepackage{graphicx,epic,eepic,epsfig,latexsym,verbatim,color}
\usepackage[inline,shortlabels]{enumitem}
\usepackage[caption=false]{subfig}
\usepackage[dvipsnames]{xcolor}
 
\usepackage{tikz}
\usetikzlibrary{chains}
\usetikzlibrary{fit}
\usetikzlibrary{arrows.meta,backgrounds,positioning}

\usepackage{epsfig}
\usetikzlibrary{shapes.symbols,patterns} %
\usepackage{pgfplots}

\usepackage[strict]{changepage}

\usepackage[marginal]{footmisc}
\usepackage{url}
\usepackage{xurl}
\usepackage{theorem}
\usepackage{thmtools}
\usepackage{nicematrix}
\usepackage{xfrac}

\newtheorem{proposition}{Proposition}

\newtheorem{theorem}[proposition]{Theorem}

\def\squareforqed{\hbox{\rlap{$\sqcap$}$\sqcup$}}
\def\qed{\ifmmode\squareforqed\else{\unskip\nobreak\hfil
\penalty50\hskip1em\null\nobreak\hfil\squareforqed
\parfillskip=0pt\finalhyphendemerits=0\endgraf}\fi}
\def\endenv{\ifmmode\;\else{\unskip\nobreak\hfil
\penalty50\hskip1em\null\nobreak\hfil\;
\parfillskip=0pt\finalhyphendemerits=0\endgraf}\fi}

\newcounter{example}

\mathchardef\ordinarycolon\mathcode`\:
\mathcode`\:=\string"8000
\def\vcentcolon{\mathrel{\mathop\ordinarycolon}}
\begingroup \catcode`\:=\active
  \lowercase{\endgroup
  \let :\vcentcolon
  }

\usepackage{hyperref}
\hypersetup{colorlinks=true,citecolor=blue,linkcolor=blue,filecolor=blue,urlcolor=blue,breaklinks=true}
\usepackage{cleveref}
\usepackage{graphicx}

\definecolor{darkblue}{RGB}{0,76,156}
\definecolor{darkkblue}{RGB}{0,0,153}
\definecolor{blue2}{RGB}{102,178,255}
\definecolor{darkred}{RGB}{195,0,0}

\RequirePackage[framemethod=default]{mdframed}
\newmdenv[skipabove=7pt,
skipbelow=7pt,
backgroundcolor=darkblue!15,
innerleftmargin=5pt,
innerrightmargin=5pt,
innertopmargin=5pt,
leftmargin=0cm,
rightmargin=0cm,
innerbottommargin=5pt,
linewidth=1pt]{tBox}

\newmdenv[skipabove=7pt,
skipbelow=7pt,
backgroundcolor=blue2!25,
innerleftmargin=5pt,
innerrightmargin=5pt,
innertopmargin=5pt,
leftmargin=0cm,
rightmargin=0cm,
innerbottommargin=5pt,
linewidth=1pt]{dBox}

\newmdenv[skipabove=7pt,
skipbelow=7pt,
backgroundcolor=darkred!15,
innerleftmargin=5pt,
innerrightmargin=5pt,
innertopmargin=5pt,
leftmargin=0cm,
rightmargin=0cm,
innerbottommargin=5pt,
linewidth=1pt]{rBox}

\newcommand{\nc}{\newcommand}
\nc{\bra}[1]{\langle#1|}
\nc{\ket}[1]{|#1\rangle}
\nc{\ketbra}[2]{\lvert#1\rangle\!\langle#2\rvert}
\nc{\braket}[2]{\langle#1|#2\rangle}

\DeclarePairedDelimiterX{\infdivx}[2]{(}{)}{%
  #1\;\delimsize\|\;#2%
}

\nc{\proj}[1]{| #1\rangle\!\langle #1 |}
\nc{\avg}[1]{\langle#1\rangle}
\nc{\smfrac}[2]{\mbox{$\frac{#1}{#2}$}}
\nc{\tr}{\operatorname{Tr}}
\nc{\ox}{\otimes}
\nc{\dg}{\dagger}
\nc{\dn}{\downarrow}
\nc{\cA}{{\cal A}}
\nc{\cB}{{\cal B}}
\nc{\cC}{{\cal C}}
\nc{\cD}{{\cal D}}
\nc{\cE}{{\cal E}}
\nc{\cF}{{\cal F}}
\nc{\cG}{{\cal G}}
\nc{\cH}{{\cal H}}
\nc{\cI}{{\cal I}}
\nc{\cJ}{{\cal J}}
\nc{\cK}{{\cal K}}
\nc{\cL}{{\cal L}}
\nc{\cM}{{\cal M}}
\nc{\cN}{{\cal N}}
\nc{\cO}{{\cal O}}
\nc{\cP}{{\cal P}}
\nc{\cQ}{{\cal Q}}
\nc{\cR}{{\cal R}}
\nc{\cS}{{\cal S}}
\nc{\cT}{{\cal T}}
\nc{\cU}{{\cal U}}
\nc{\cV}{{\cal V}}
\nc{\cX}{{\cal X}}
\nc{\cY}{{\cal Y}}
\nc{\cZ}{{\cal Z}}
\nc{\cW}{{\cal W}}
\nc{\csupp}{{\operatorname{csupp}}}
\nc{\qsupp}{{\operatorname{qsupp}}}
\nc{\var}{{\operatorname{var}}}
\nc{\rar}{\rightarrow}
\nc{\lrar}{\longrightarrow}
\nc{\polylog}{{\operatorname{polylog}}}
\nc{\wt}{{\operatorname{wt}}}
\nc{\supp}{{\operatorname{supp}}}

\nc{\argmin}{{\operatorname{argmin}}}

\makeatletter
\newcommand{\tpmod}[1]{{\@displayfalse\pmod{#1}}}
\makeatother

\def\x{\xi}

\nc{\RR}{{{\mathbb R}}}
\nc{\CC}{{{\mathbb C}}}
\nc{\FF}{{{\mathbb F}}}
\nc{\NN}{{{\mathbb N}}}
\nc{\ZZ}{{{\mathbb Z}}}
\nc{\PP}{{{\mathbb P}}}
\nc{\QQ}{{{\mathbb Q}}}
\nc{\UU}{{{\mathbb U}}}
\nc{\EE}{{{\mathbb E}}}
\nc{\id}{{\operatorname{id}}}

\nc{\CHSH}{{\operatorname{CHSH}}}

\nc{\rU}{\mbox{U}}

\nc{\ob}[1]{#1}

\nc{\SEP}{{\text{\rm SEP}}}
\nc{\NS}{{\text{\rm NS}}}
\nc{\LOCC}{{\text{\rm LOCC}}}
\nc{\PPT}{{\text{\rm PPT}}}
\nc{\EXT}{{\text{\rm EXT}}}
\nc{\Sym}{{\operatorname{Sym}}}

\nc{\ERLO}{{E_{\text{r,LO}}}}
\nc{\ERLOCC}{{E_{\text{r,LOCC}}}}
\nc{\ERPPT}{{E_{\text{r,PPT}}}}
\nc{\ERLOCCinfty}{{E^{\infty}_{\text{r,LOCC}}}}
\nc{\Aram}{{\operatorname{\sf A}}}
\newtheorem{problem}{Problem}

\makeatletter
\def\grd@save@target#1{%
  \def\grd@target{#1}}
\def\grd@save@start#1{%
  \def\grd@start{#1}}
\tikzset{
  grid with coordinates/.style={
    to path={%
      \pgfextra{%
        \edef\grd@@target{(\tikztotarget)}%
        \tikz@scan@one@point\grd@save@target\grd@@target\relax
        \edef\grd@@start{(\tikztostart)}%
        \tikz@scan@one@point\grd@save@start\grd@@start\relax
        \draw[minor help lines,magenta] (\tikztostart) grid (\tikztotarget);
        \draw[major help lines] (\tikztostart) grid (\tikztotarget);
        \grd@start
        \pgfmathsetmacro{\grd@xa}{\the\pgf@x/1cm}
        \pgfmathsetmacro{\grd@ya}{\the\pgf@y/1cm}
        \grd@target
        \pgfmathsetmacro{\grd@xb}{\the\pgf@x/1cm}
        \pgfmathsetmacro{\grd@yb}{\the\pgf@y/1cm}
        \pgfmathsetmacro{\grd@xc}{\grd@xa + \pgfkeysvalueof{/tikz/grid with coordinates/major step}}
        \pgfmathsetmacro{\grd@yc}{\grd@ya + \pgfkeysvalueof{/tikz/grid with coordinates/major step}}
        \foreach \x in {\grd@xa,\grd@xc,...,\grd@xb}
        \node[anchor=north] at (\x,\grd@ya) {\pgfmathprintnumber{\x}};
        \foreach \y in {\grd@ya,\grd@yc,...,\grd@yb}
        \node[anchor=east] at (\grd@xa,\y) {\pgfmathprintnumber{\y}};
      }
    }
  },
  minor help lines/.style={
    help lines,
    step=\pgfkeysvalueof{/tikz/grid with coordinates/minor step}
  },
  major help lines/.style={
    help lines,
    line width=\pgfkeysvalueof{/tikz/grid with coordinates/major line width},
    step=\pgfkeysvalueof{/tikz/grid with coordinates/major step}
  },
  grid with coordinates/.cd,
  minor step/.initial=.2,
  major step/.initial=1,
  major line width/.initial=2pt,
}
\makeatother

\usepackage{thmtools}
\usepackage{thm-restate}
\usepackage{etoolbox}
\makeatletter
\def\problem@s{}
\newcounter{problems@cnt}

\newcommand{\allproblems}{\problem@s}
\makeatother

\usepackage{authblk} %
\usepackage[a4paper,top=20mm,bottom=23mm,left=22mm,right=22mm,headheight=15pt,footskip=11mm]{geometry}

\usepackage[utf8]{inputenc}
\usepackage[T1]{fontenc}
\usepackage{lmodern}
\usepackage[scaled=0.98]{zi4}
\usepackage{microtype}
\usepackage{listings}
\usepackage[most]{tcolorbox}
\usepackage{varwidth}
\usepackage{fancyhdr}
\usepackage{titlesec}
\pgfplotsset{compat=1.18}
\usepackage[numbers]{natbib}
\usepackage{multirow}
\usepackage{booktabs}
\usepackage{array}
\usepackage{cleveref}
\usepackage{tabularx}
\usepackage{soul}
\usepackage[export]{adjustbox}

\definecolor{colortwo}{rgb}{0.4,0.77,0.17}
\definecolor{colorthree}{rgb}{0.01,0.51,0.93}
\definecolor{darkgray}{rgb}{0.3,0.3,0.3}

\usepackage{tikz}
\usetikzlibrary{quantikz2}

\NiceMatrixOptions{cell-space-limits = 1pt}

\definecolor{QSTInk}{RGB}{31,40,54}
\definecolor{QSTNavy}{RGB}{24,55,91}
\definecolor{QSTMuted}{RGB}{102,116,135}
\definecolor{QSTPanelBlueDark}{RGB}{21,64,107}
\definecolor{QSTPanelBlueFrame}{RGB}{92,139,181}
\definecolor{QudeBlue}{RGB}{34,55,199}
\definecolor{QudeViolet}{RGB}{108,49,225}
\definecolor{QITLeanDecl}{RGB}{15,77,151}
\definecolor{QITLeanNamespace}{RGB}{16,108,137}
\definecolor{QITLeanProof}{RGB}{49,118,89}
\definecolor{QITLeanType}{RGB}{36,97,153}
\definecolor{QITLeanDomain}{RGB}{38,83,128}
\definecolor{QITLeanComment}{RGB}{94,111,124}
\definecolor{QITLeanString}{RGB}{26,118,116}
\definecolor{QITLeanWarn}{RGB}{145,69,60}

\DeclareRobustCommand{\LeanQuantumAlg}{%
  \textnormal{\textsc{Lean-QuantumAlg}}%
}
\newcommand{\QudeLeapLogo}[1]{%
  \includegraphics[#1]{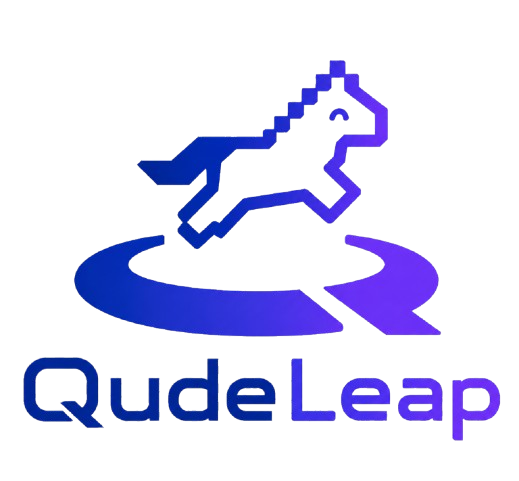}%
}

\titleformat{\section}
  {\Large\sffamily\bfseries\color{QSTPanelBlueDark}}
  {\thesection}{0.65em}{}
  [\vspace{0.15em}\color{QSTPanelBlueFrame!55}\titlerule]
\titleformat{\subsection}
  {\large\sffamily\bfseries\color{QSTNavy}}
  {\thesubsection}{0.65em}{}
\titleformat{\subsubsection}
  {\normalsize\sffamily\bfseries\color{QSTInk}}
  {\thesubsubsection}{0.65em}{}

\makeatletter
\renewcommand{\maketitle}{%
  \thispagestyle{empty}%
  \begingroup
  \setlength{\parindent}{0pt}%
  \noindent
  \begin{minipage}[t]{0.76\textwidth}
    \vspace{0pt}%
    {\raggedright\hyphenpenalty=10000\exhyphenpenalty=10000
      \sffamily\bfseries\color{QSTInk}\fontsize{23.5}{27.5}\selectfont
      \@title\par}%
  \end{minipage}%
  \hfill
  \begin{minipage}[t]{0.18\textwidth}
    \vspace{0pt}\raggedleft
    \QudeLeapLogo{width=2.65cm}%
  \end{minipage}
  \vspace{7mm}

  \noindent
  {\color{QudeBlue}\rule{0.22\textwidth}{1.8pt}}%
  {\color{QudeViolet}\rule{0.78\textwidth}{1.8pt}}\par
  \vspace{5mm}

  \begin{center}
    {\sffamily\normalsize\color{QSTInk}\@author\par}%
  \end{center}
  \vspace{3mm}
  \@thanks
  \endgroup
  \setcounter{footnote}{0}%
}
\makeatother

\lstdefinelanguage{Lean4}{
  sensitive=true,
  alsoletter={_},
  morekeywords=[1]{
    abbrev,class,def,deriving,example,inductive,instance,lemma,
    noncomputable,opaque,structure,theorem
  },
  morekeywords=[2]{
    end,export,namespace,open,private,protected,public,section,variable,
    variables,where
  },
  morekeywords=[3]{
    by,calc,change,do,else,exact,exists,for,forall,from,fun,have,if,
    inferInstance,infix,infixl,infixr,intro,intros,let,letI,macro,match,
    obtain,rcases,refine,rfl,rw,show,syntax,termination_by,then,with
  },
  morekeywords=[4]{
    Bool,False,Fin,Int,List,Nat,None,Option,Prop,Real,Set,Some,Sort,
    String,True,Type
  },
  morekeywords=[5]{axiom,sorry,admit},
  morekeywords=[6]{
    Circuit,Data,InputClass,P256DomainParameters,PublicEndpointWitness,
    PublicTheoremShape,PureState,ResourceParameters,ScalarRecoveryInterface,
    StateVector,ZMod
  },
  morecomment=[l]{--},
  morecomment=[s]{/-}{-/},
  morestring=[b]",
}

\lstdefinestyle{QudeLeanListing}{
  language=Lean4,
  basicstyle=\fontsize{9.8}{11.5}\selectfont\ttfamily\color{QSTInk},
  keywordstyle=[1]\color{QITLeanDecl}\bfseries,
  keywordstyle=[2]\color{QITLeanNamespace}\bfseries,
  keywordstyle=[3]\color{QITLeanProof},
  keywordstyle=[4]\color{QITLeanType}\bfseries,
  keywordstyle=[5]\color{QITLeanWarn}\bfseries,
  keywordstyle=[6]\color{QITLeanDomain}\bfseries,
  commentstyle=\color{QITLeanComment},
  stringstyle=\color{QITLeanString},
  numbers=none,
  xleftmargin=0pt,
  framexleftmargin=0pt,
  columns=fullflexible,
  keepspaces=true,
  showstringspaces=false,
  tabsize=2,
  breaklines=true,
  breakatwhitespace=false,
  breakindent=1.15em,
  breakautoindent=true,
  aboveskip=0pt,
  belowskip=0pt,
  literate=
    {_}{{\textunderscore\allowbreak}}1
    {∀}{{\ensuremath{\forall}}}1
    {∃}{{\ensuremath{\exists}}}1
    {→}{{\ensuremath{\to}}}1
    {←}{{\ensuremath{\leftarrow}}}1
    {↔}{{\ensuremath{\leftrightarrow}}}1
    {∧}{{\ensuremath{\wedge}}}1
    {∨}{{\ensuremath{\vee}}}1
    {¬}{{\ensuremath{\neg}}}1
    {≤}{{\ensuremath{\leq}}}1
    {≥}{{\ensuremath{\geq}}}1
    {≠}{{\ensuremath{\neq}}}1
    {⟨}{{\ensuremath{\langle}}}1
    {⟩}{{\ensuremath{\rangle}}}1
    {λ}{{\ensuremath{\lambda}}}1
    {α}{{\ensuremath{\alpha}}}1
    {β}{{\ensuremath{\beta}}}1
    {θ}{{\ensuremath{\theta}}}1
    {ℕ}{{\ensuremath{\mathbb{N}}}}1
    {ℤ}{{\ensuremath{\mathbb{Z}}}}1
    {ℚ}{{\ensuremath{\mathbb{Q}}}}1
    {ℝ}{{\ensuremath{\mathbb{R}}}}1
    {×}{{\ensuremath{\times}}}1
    {ˣ}{{\ensuremath{^{\times}}}}1
    {⁻¹}{{\ensuremath{^{-1}}}}2
}

\newtcblisting[auto counter]{leanlisting}[3][]{%
  enhanced,
  breakable,
  listing only,
  listing engine=listings,
  listing options={style=QudeLeanListing},
  colback=gray!5,
  colframe=gray!55,
  boxrule=0.4pt,
  arc=0.6mm,
  left=4pt,
  right=4pt,
  top=4pt,
  bottom=4pt,
  title={Listing~\thetcbcounter: #2},
  fonttitle=\sffamily\small,
  coltitle=black,
  attach boxed title to top left={yshift=-2mm,xshift=4mm},
  varwidth boxed title=\dimexpr\linewidth-8mm\relax,
  boxed title style={colback=gray!18,sharp corners,boxrule=0pt,boxsep=2pt},
  label={#3},
  before skip=0.6\baselineskip,
  after skip=0.65\baselineskip,
  #1
}

\usepackage{soul}
\soulregister\cite7
\soulregister\ref7
\soulregister\pageref7

\NewDocumentCommand{\ddx}{o m}{%
  \IfNoValueTF{#1}
  {\frac{\operatorname{d}}{\operatorname{d}\!#2}}
  {\frac{\operatorname{d}^{\,#1}}{\operatorname{d}\!#2^{\,#1}}}%
}

\usepackage{pifont}
\newcommand{\cmark}{\ding{51}}
\newcommand{\pmark}{\ensuremath{-}}
\newcommand{\xmark}{\ding{55}}

\newcommand{\specialcell}[2][c]{%
  \begin{tabular}[#1]{@{}c@{}}#2\end{tabular}}

\title{\huge Building Shor's Algorithm in Lean: \\ An Agentic Formalization of Quantum \\ Attacks on RSA-2048 and P-256}

\author[1,2]{Lei Zhang\thanks{leizhang116.4@gmail.com}}
\author[1]{Yusheng Zhao}
\author[1,2]{Hongshun Yao}
\author[2]{Xin Wang\thanks{felixxinwang@hkust-gz.edu.cn}}
\affil[1]{\small QudeLeap Research, Shanghai 200030, China}
\affil[2]{\small The Hong Kong University of Science and Technology (Guangzhou), Guangdong 511453, China}
\date{}

\begin{document}
\maketitle

\vspace{-5mm}
\begin{center}
  \small\sffamily\color{QSTMuted}%
  \raisebox{-0.5ex}{\QudeLeapLogo{height=3.2mm}}\hspace{2pt}%
  \href{https://github.com/QudeLeap/Lean-QuantumAlg}{github.com/QudeLeap/Lean-QuantumAlg}%
\end{center}

\begin{abstract}
Large language models are increasingly assisting with demanding formal theorem-proving tasks, particularly when grounded in machine-checked libraries such as Lean. Agentic systems further amplify this process by searching, reusing, and extending existing formal developments to uncover new discoveries. In quantum computing, Shor's algorithm and its variants present such a demanding case for Lean formalization. In this work, we formalize this algorithm family in Lean through agentic formalization: software agents analyze sources, write Lean code and repair proofs, with human review of the scientific claims and machine checking of the resulting formal proofs. Our formalization develops the mathematical foundations for analyzing quantum attacks in two cryptographic settings: a 2048-bit modulus in the RSA-2048 and the standardized elliptic curve over a 256-bit prime field (P-256). To support these analyses, the formalization ranges from quantum algorithms for order finding to reversible quantum circuits for modular and elliptic-curve arithmetic. Based on [Quantum 5, 433] and [ASIACRYPT 2017, 241--270], we formalize the logical resource estimates for RSA-2048 and P-256, respectively, and provide additional estimates of classical operations. We expect the results pave the way for broader machine-checked quantum cryptanalysis and represent a step toward AI-assisted design and verification of quantum algorithms.
\end{abstract}

\section{Introduction}

Formalizing quantum algorithms is important because their application-level claims connect mathematical guarantees, circuit behavior, success probabilities, classical post-processing, and resource estimates. The field spans algebraic algorithms and search~\cite{childs2021lectures,dewolf2019lecturenotes,montanaro2016quantum}, as well as quantum simulation and scientific computation~\cite{lin2022qasc}. In each area, abstract arguments interact with choices of representation, subroutines, measurement, and repetition that determine what an algorithm returns and how its cost scales. A machine-checked development records these links as explicit definitions and theorems, giving researchers a precise basis for comparing constructions and extending them.

Shor's algorithm~\cite{shor1995factorization} is a demanding test case for this proof chain because it gives polynomial-time quantum procedures for integer factorization and discrete logarithms. Factorization underlies the Rivest--Shamir--Adleman (RSA) cryptosystem~\cite{rivest1978method}; factoring a 2048-bit public modulus exposes the prime factors needed to derive the RSA-2048 private key. The elliptic-curve discrete-logarithm problem asks for the private scalar that maps a public base point to a public key~\cite{roetteler2017ecdlp}. P-256~\cite{nist2023sp800186} is the 256-bit prime-field curve standardized by the National Institute of Standards and Technology. Both problems admit a hidden-subgroup interpretation~\cite{jozsa2001HiddenSubgroup,childsVanDam2010AlgebraicProblems}. Their cryptographic instances connect quantum sampling with number theory, reversible arithmetic, classical recovery, probability analysis, and resource accounting.

Existing formal verification results establish increasingly broad portions of Shor's algorithm. Ref.~\cite{chareton2021automated} presents Qbricks and uses it to verify a parameterized order-finding circuit, a success bound, and asymptotic circuit resources. Ref.~\cite{feng2021quantum} verifies a hybrid factorization program with classical and quantum variables, including a lower bound on factor-recovery success, while representing controlled modular exponentiation as an abstract unitary. Ref.~\cite{li2024qafny} presents Qafny and verifies Shor order finding, continued-fraction post-processing, and its probability calculation under a specified modular-multiplication behavior. Ref.~\cite{peng2023formally} certifies a factoring implementation that connects reversible modular multiplication, quantum phase estimation, measurement, continued fractions, classical factor recovery, success probability, and gate counts.

These developments also motivate a reusable mathematical layer for studying new quantum algorithms and variants. For Shor's algorithm, that layer includes reductions from cryptographic problems, distributions of quantum measurement outcomes, probability bounds for recoverable events, reversible arithmetic, classical recovery, and resource formulas. Stable names and module locations make the assumptions and conclusions of each component visible. Researchers can then retrieve an intermediate theorem, inspect its dependencies, compare related formulations, and assemble the checked components into another analysis while retaining a trace from the final claim to its mathematical premises.

Lean provides a common language for executable definitions, mathematical structures, and checked proofs. Its libraries place general mathematics and computer-science foundations in Mathlib~\cite{mathlib} and CSLib~\cite{cslib}. Lean-QIT~\cite{leanqit} extends this ecosystem with finite-dimensional quantum information theory, supplying a foundation for formal quantum developments. LeanDojo~\cite{yang2023leandojo} gives agents programmatic access to Lean repositories and proof states. Together, checked libraries and repository interfaces allow agents to locate relevant statements, retrieve their assumptions, and reuse verified components. This combination supports a quantum-algorithm library in which mathematical arguments, circuit properties, probability claims, and resource formulas share one proof language.

In this work, we formalize mathematical reductions and selected quantum circuit components used to analyze RSA-2048 factorization and P-256 scalar recovery in Lean. The RSA branch develops order finding, factor recovery, reversible modular arithmetic, and the Eker\aa--H\aa stad route~\cite{ekera2017quantum}. The P-256 branch develops discrete logarithms in finite cyclic groups, prime-field arithmetic, elliptic-curve operations, and scalar recovery. These branches expose reusable statements for number theory, probability, reversible circuits, classical post-processing, and resource accounting. To the best of our knowledge, this is the first Lean development to give a machine-checked P-256 scalar-recovery statement with a quantitative success bound and logical resource values. We constructed the library through agentic formalization~\cite{jing2026agentic}: agents researched sources, decomposed claims, wrote Lean code, and repaired proofs; human review covered scientific statements and source alignment; and machine checking verified the resulting formal proofs. The resulting organization provides a Lean foundation for further machine-checked studies of quantum algorithms and their cryptographic applications.
\section{Main Results}
\label{sec:main-results}

This section presents the two application-level results of the formalization. The first concerns factorization of the public modulus used by the 2048-bit Rivest--Shamir--Adleman system, called RSA-2048. The second concerns recovery of the private scalar associated with a public point on P-256. In each case, we begin with the public cryptographic data, formulate the computational problem, and follow the quantum and classical stages that lead to the requested output. The resulting theorem states the success guarantee and resource profile, and the accompanying table places that result beside earlier analyses. The later sections on formalizing RSA-2048 and P-256 develop the number-theoretic reductions, reversible arithmetic, and recovery procedures used along these two routes.

Both results use a logical-circuit resource model. The resource profiles count logical qubits, Toffoli gates, and the depth quantity reported by the corresponding source. For P-256, Ref.~\cite{roetteler2017ecdlp} reports maximal Toffoli-gate depth. For one RSA-2048 run, Ref.~\cite{gidney2019factor} reports measurement depth. Circuit noise, quantum error correction, scheduling, and physical resources require additional models beyond these logical profiles. Each success probability refers to recovery of the output specified by the relevant problem, and each classical-operation count records the abstract arithmetic steps in the associated recovery procedure.

\subsection{RSA-2048}

The RSA cryptosystem~\cite{rivest1978method} publishes a modulus $N=pq$ formed from two secret primes $p$ and $q$. RSA-2048 fixes the binary length of $N$ to 2048 bits. The attacker receives $N$ and seeks a value $d$ equal to either $p$ or $q$; division by $d$ then gives the other prime. Recovering this factorization provides the secret number-theoretic data needed to derive the corresponding private key. The application-level input and output are therefore captured by the following problem.

\begin{problem}[Integer Factorization]\label{prob:factorization}
Let $p$ and $q$ be unknown distinct primes, and let $N=pq$ be known.
Given $N$, find $d\in\{p,q\}$.
\end{problem}

Problem~\ref{prob:factorization} gives the finish line for the algorithm: one nontrivial factor of $N$. Shor's factoring algorithm~\cite{shor1995factorization} reaches this output through order finding. For a selected residue $a$ coprime to $N$, the quantum stage determines the period of repeated multiplication by $a$ modulo $N$. A suitable even period makes the powers immediately before the return to $1$ reveal two candidates whose greatest common divisors with $N$ yield the prime factors. Reversible modular addition, multiplication, and exponentiation implement the periodic action coherently, and the classical stage converts a recoverable measurement result into the requested value $d$.

One factoring attempt follows this reduction as a sequence of transformations. A superposition of exponents controls modular exponentiation, so exponents separated by the unknown period produce the same modular value. Measuring the quantum registers then provides numerical information from which continued fractions reconstruct a candidate period. The classical stage validates that candidate by checking the modular power relation and tests the associated greatest common divisors. A recoverable candidate yields $d$; another outcome starts a new attempt. This data flow links the public modulus to a periodic circuit, the circuit to a candidate order, and the candidate order to a prime factor.

The formalization follows two routes to this common factoring goal. The textbook Shor route supplies the order-finding and factor-recovery analysis used to express success under repetition. The Eker\aa--H\aa stad short-discrete-logarithm construction analyzed in Ref.~\cite{gidney2019factor} supplies the concrete RSA-2048 logical-circuit baseline. One run uses $6190$ live logical qubits, $2.7\times10^9$ Toffoli gates, and $2.14\times10^9$ measurement depth, and its classical post-processing recovers the factors with probability above $99\%$ conditional on a correct quantum run. The three-trial calculation assigns each factoring attempt failure probability at most one half. Independence then reduces the joint failure probability to at most $1/8$, which meets the target of at most $1/3$. Sequential execution preserves the live-qubit footprint and triples both the Toffoli count and the reported depth.

The resulting success event is the return of a value $d\in\{p,q\}$. The repeated-trial calculation gives probability at least $2/3$. Applying the one-run values reported in Ref.~\cite{gidney2019factor} to the same three attempts preserves the peak logical-qubit count and triples the Toffoli count and depth. The recovery bookkeeping adds an exact count of abstract classical arithmetic operations. Theorem~\ref{thm:rsa2048} collects the resulting values.

\begin{theorem}[RSA-2048]\label{thm:rsa2048}
Suppose an RSA-2048 public key has known public modulus $N$, whose private prime factors are unknown distinct primes $p$ and $q$. There exists a quantum algorithm that returns $d\in\{p,q\}$ with probability at least $2/3$, using $6.19\times 10^3$ logical qubits, $8.1\times 10^9$ Toffoli gates, $6.42\times 10^9$ maximal circuit depth, and $3.69\times 10^4$ classical arithmetic operations.
\end{theorem}

Theorem~\ref{thm:rsa2048} combines the cryptographic output in Problem~\ref{prob:factorization} with an explicit success threshold and a concrete resource profile. Its quantum resources describe three sequential attempts, and its classical count describes the arithmetic recovery steps recorded for those attempts. The later section on formalizing RSA-2048 starts from the period-finding reduction, builds the reversible modular-arithmetic path, and then presents the Shor and Eker\aa--H\aa stad recovery procedures that support this application-level statement.

\begin{table}[t]
\centering
\setlength{\tabcolsep}{1em}
\caption{A comparison of some results that estimate the logical resources required to solve the integer-factorization problem for RSA-2048, without considering circuit noise. Here $n$ denotes the bit length of the RSA-2048 modulus.}~\label{tab:rsa}
\resizebox{\linewidth}{!}{
\begin{tabular}{p{11em} p{17em} p{12em}}
\toprule
Result & Logical resources & Success probability \\
\midrule
\addlinespace
{Asymptotic analysis~\cite{shor1995factorization}} & {$O(n)$ qubits \par $O(n^3)$ elementary gates} & {High probability after repetition} \\
\addlinespace
{Gate-level estimate~\cite{gidney2019factor}} & {$6.19\times 10^3$ qubits \par $2.7\times 10^9$ Toffoli gates \par $2.14\times 10^9$ measurement depth per run} & {$\geq 99\%$ classical recovery conditional on a correct quantum run} \\
\addlinespace
{Lean-formalized \par (Ours; Theorem~\ref{thm:rsa2048})} & {$6.19\times 10^3$ qubits \par $8.1\times 10^9$ Toffoli gates \par $6.42\times 10^9$ maximal circuit depth \par $3.69\times 10^4$ classical arithmetic operations} & $\geq 2/3$ \\
\addlinespace
\bottomrule
\end{tabular}
}
\end{table}

Table~\ref{tab:rsa} shows the progression from an asymptotic factoring algorithm to an RSA-2048 gate-level estimate and then to the application statement in Theorem~\ref{thm:rsa2048}. The first two rows establish the algorithmic scaling and the concrete one-run circuit baseline. The final row adds an explicit repeated-run success threshold, the corresponding aggregate logical resources, and a classical recovery count in one machine-checked result.

\subsection{P-256}

P-256~\cite{nist2023sp800186} is a standardized elliptic curve $E$ over a 256-bit prime field $\mathbb{F}_p$. Its public domain parameters include a base point $P$ that generates a subgroup of prime order $r$. A private scalar $m\in\mathbb{Z}_r$ determines the public point $Q=[m]P$, where $[m]P$ denotes scalar multiplication on the curve. The attacker receives the curve parameters, $P$, $Q$, and $r$, and seeks the scalar $m$ that links the two public points. This input-output relation gives the P-256 instance of the elliptic-curve discrete-logarithm problem.

\begin{problem}[Elliptic-Curve Discrete Logarithm]\label{prob:ecdlp}
Let $E/\mathbb{F}_p$ be an elliptic curve, let $P\in E(\mathbb{F}_p)$
have prime order $r$, and let $Q=[m]P$ for an unknown
$m\in\mathbb{Z}_r$. Given $(E,p,P,Q,r)$, find $m$.
\end{problem}

Problem~\ref{prob:ecdlp} asks the algorithm to invert the scalar action that produced $Q$. The elliptic-curve construction described in Ref.~\cite{proos2004shor} begins by placing pairs of candidate scalars in superposition and coherently evaluating the corresponding combinations of $P$ and $Q$. Different pairs can produce the same curve point because $Q$ already contains the hidden multiplier $m$. Quantum interference concentrates the subsequent measurement outcomes on relations determined by this multiplier. Classical post-processing reads a recoverable relation and solves for $m$, turning information exposed by the quantum stage into the output required by the problem.

The useful measurement information comes from entire families of scalar pairs that reach the same curve point. Coherent point evaluation preserves the relation among these pairs until the point register is measured. The two scalar registers then retain a shared dependence on the hidden multiplier, and their interference makes this dependence visible as correlated measurement values. Classical recovery checks whether the observed relation can be solved and performs the required modular arithmetic when it can. The algorithm therefore moves from public curve data to a coherent group action, from that action to correlated measurement values, and from those values to the private scalar.

The coherent group evaluation becomes a circuit through three layers of arithmetic. Reversible operations in $\mathbb{F}_p$ implement the additions, multiplications, squarings, and inversions used by affine curve formulas. These field operations compose into controlled point additions, and schedules of controlled additions implement scalar multiplication from the two superposed index registers. Ref.~\cite{roetteler2017ecdlp} analyzes this gate-level construction for P-256 and reports $2330$ logical qubits, $1.26\times10^{11}$ Toffoli gates, and maximal Toffoli-gate depth $1.16\times10^{11}$. After a recoverable measurement outcome is obtained, the direct scalar-recovery bookkeeping uses seven abstract classical operations.

For the P-256 result, success means that the returned value equals the scalar $m$ in the public relation $Q=[m]P$. The quantitative profile assigns this event probability at least $2/3$ and records one run of the logical circuit together with the direct classical recovery. The following theorem packages that application-level output, probability, and resource statement.

\begin{theorem}[P-256]\label{thm:p256}
Suppose a P-256 instance has known base point $P$ of prime order $r$ and known public key point $Q=[m]P$ for unknown private scalar $m$. There exists a quantum algorithm that returns $m$ with probability at least $2/3$, using $2.33\times 10^3$ logical qubits, $1.26\times 10^{11}$ Toffoli gates, $1.16\times 10^{11}$ maximal Toffoli-gate depth, and $7$ classical operations.
\end{theorem}

Theorem~\ref{thm:p256} connects the public-point relation in Problem~\ref{prob:ecdlp} to a quantitative scalar-recovery guarantee. Its resource tuple follows the reversible circuit from prime-field arithmetic through controlled curve operations, and its classical count begins when the quantum measurement has supplied a recoverable relation. The later section on formalizing P-256 develops these layers in order: the generic discrete-logarithm argument, reversible prime-field arithmetic, elliptic-curve point operations, scalar multiplication, and final recovery.

\begin{table}[t]
\centering
\setlength{\tabcolsep}{1em}
\caption{A comparison of some results that estimate the resources required to solve the elliptic-curve discrete-logarithm problem for P-256, without considering circuit noise.}~\label{tab:ecc}
\resizebox{\linewidth}{!}{
\begin{tabular}{p{11em} p{17em} p{12em}}
\toprule
Result & Logical resources & Success probability \\
\midrule
\addlinespace
{Approximate analysis \cite{proos2004shor}} & {$\approx1.5\times10^3$ qubits \par $\approx1.81\times10^{10}$ Toffoli gates} & {Close to $1$} \\
\addlinespace
{Gate-level estimate \cite{roetteler2017ecdlp}} & {$2.33\times 10^3$ qubits \par $1.26\times 10^{11}$ Toffoli gates \par $1.16\times 10^{11}$ maximal Toffoli-gate depth} & {Close to $1$} \\
\addlinespace
{Lean-formalized \par (Ours; Theorem~\ref{thm:p256})} & {$2.33\times 10^3$ qubits \par $1.26\times 10^{11}$ Toffoli gates \par $1.16\times 10^{11}$ maximal Toffoli-gate depth \par $7$ classical operations} & $\geq 2/3$ \\
\addlinespace
\bottomrule
\end{tabular}
}
\end{table}

Table~\ref{tab:ecc} moves from an approximate elliptic-curve construction to a simulated gate-level P-256 estimate and finally to Theorem~\ref{thm:p256}. The comparison shows how success and resource accounting become progressively more explicit across the three levels. The final row collects the concrete logical-circuit profile, the stated recovery probability, and the direct classical-operation count in the formalized application result.
\section{Formalizing Quantum Algorithm Proofs}

\subsection{Agentic Formalization}

Agentic formalization is a workflow in which software agents conduct substantive source research and proof development, human review examines the mathematical claims and their alignment with cited sources, and machine checking verifies the resulting formal proofs. Software agents may propose definitions, theorem statements, proofs, and revisions, while authors decide which claims and changes enter the project. Machine checking establishes that a proof has the stated Lean type under the imported definitions and assumptions. We use the term only for this supervised division of work.

The workflow begins when authors select a scientific question and the sources that should support it. Software agents retrieve the relevant source passages, separate a claim into mathematical and computational obligations, and construct Lean declarations for those obligations. The Lean kernel then checks the proof terms. Authors inspect the formal assumptions, compare the checked statement with the source, and review the corresponding manuscript language before accepting a change. The accepted declarations, source locations, review decisions, and manuscript claims form the artifacts passed to the next stage. Figure~\ref{fig:agentic-workflow} in Appendix~\ref{app:workflow} summarizes this sequence, marks the trust boundary, and assigns responsibility for each artifact.

Stable names and module locations make a formal library usable by software agents. An agent can locate a statement, retrieve its dependencies and assumptions, compare its conclusion with a cited claim, and reuse the checked result in a later construction. This organization also gives reviewers a path from manuscript prose to a precise theorem and from that theorem to the definitions on which it depends. Appendix~\ref{app:lean-formalization} maps the RSA-2048 and P-256 claims to their final theorems and states each theorem's assumptions and conclusion.

\subsection{Formal Verification Results}

Table~\ref{tab:formulization} compares selected formal developments of Shor's factoring algorithm. The first column identifies the development. The remaining columns ask whether the formal scope reaches factor recovery, whether the development proves a quantitative success bound for the endpoint it formalizes, whether it analyzes quantum and classical resources, and whether it constructs the quantum computation at gate level. The oracle column records an implementation boundary: it names any arithmetic operation left as an abstract oracle and reports ``None'' when the arithmetic is constructed. The check marks describe properties established inside the formal development. A check mark denotes complete coverage of the stated field, a partial mark denotes coverage of part of that field, and a cross denotes absent coverage.

In the order shown, Ref.~\cite{chareton2021automated} presents Qbricks and uses it to construct and verify the order-finding circuit, prove a lower bound on the phase measurement probability, and derive symbolic gate and width formulas. The Qbricks formal result ends before continued fractions and factor recovery, and it constructs the required modular arithmetic. Ref.~\cite{feng2021quantum} verifies the hybrid control flow and factor recovery success using quantum Hoare logic~\cite{ying2011floyd} for programs with classical and quantum variables, with controlled modular exponentiation represented as an abstract unitary. Ref.~\cite{ying2026practical} extends this program logic with quantum arrays and parameterized gates. Ref.~\cite{li2024qafny} presents Qafny and uses it to verify the order-finding state evolution, continued-fraction post-processing, and probability calculation, while modular multiplication remains at an abstract arithmetic boundary in the verified Shor chain. Coq~\cite{coqDevelopmentTeam2024} provides the proof-assistant setting for the final row among the prior developments, and CoqQ~\cite{zhou2023coqq} supplies a deeply embedded quantum programming language with a machine-checked soundness proof. Ref.~\cite{peng2023formally} constructs a reversible modular multiplier that overwrites its input register. A verified generic control transformation supplies the controlled instances used by phase estimation, so the oracle column records ``None.'' The Coq development connects measurement, continued fractions, factor recovery, repetition, and quantum gate bounds. Its classical resource analysis bounds continued-fraction iterations and repeated trials. The Lean row has declarations checked by the Lean kernel in every table category and constructs modular arithmetic circuits, so its oracle column records ``None.'' The checks in this row record category coverage. Local arithmetic results connect gate-level behavior and resource counts for the same circuit. At the terminal RSA-2048 statement, the success bound, the repository count of classical arithmetic steps, and the quantum resource tuple reported in Ref.~\cite{gidney2019factor} enter through separate certificates. A future integration step will derive all terminal fields from a circuit execution whose input consists only of the public modulus.

Our organizational target is a library in which mathematical reductions, probability bounds, reversible circuits, classical recovery, and resource statements can be located and reused independently. The RSA-2048 branch provides declarations in each of these categories and makes their current points of assembly explicit. The same organization extends beyond factoring: the P-256 branch combines discrete-logarithm recovery in a finite cyclic group with prime-field arithmetic, elliptic-curve operations, and a logical resource statement. The next two sections present these branches from their computational problems through their quantitative application results.

\begin{table}[t]
\centering
\setlength{\tabcolsep}{1em}
\caption{A comparison of some formal verification results for Shor's factoring algorithm. Here ``Oracle assumption'' lists an arithmetic oracle left unimplemented; \cmark and \pmark indicate complete and partial coverage, respectively; \xmark{} indicates a property not established.}~\label{tab:formulization}
\resizebox{\linewidth}{!}{
\begin{tabular}{lcccccc}
\toprule
Formal development & \specialcell{End-to-end} & \specialcell{Success\\probability} & \specialcell{Quantum\\resource} & \specialcell{Classical\\resource} & \specialcell{Gate-level\\circuit} & \specialcell{Oracle assumption} \\
\midrule
\addlinespace
Qbricks~\cite{chareton2021automated} & \xmark & \cmark & \cmark & \xmark & \cmark & None \\
\addlinespace
Hoare logic~\cite{feng2021quantum} & \pmark & \cmark & \xmark & \xmark & \xmark & Modular exponentiation \\
\addlinespace
Qafny~\cite{li2024qafny} & \xmark & \cmark & \xmark & \xmark & \xmark & Modular multiplication \\
\addlinespace
Coq~\cite{peng2023formally} & \cmark & \cmark & \cmark & \pmark & \cmark & None \\
\addlinespace
Lean (Ours) & \cmark & \cmark & \cmark & \cmark & \cmark & None \\
\addlinespace
\bottomrule
\end{tabular}
}
\end{table}
\section{Formalizing RSA-2048}

\subsection{Order Finding}

Write $N=pq$ for the modulus in Problem~\ref{prob:factorization}, and choose a residue $a$ with $1<a<N$. Computing $\gcd(a,N)$ already returns a factor when the greatest common divisor exceeds one. Otherwise, $a$ is a unit modulo $N$, and its multiplicative order $r$ is the least positive integer satisfying $a^r\equiv1\pmod N$. Shor's reduction~\cite{shor1995factorization} asks the quantum subroutine to recover this period. The reduction thereby replaces an unstructured search for $p$ or $q$ with the periodic action of multiplication by $a$ on residues modulo $N$.

Suppose that $r$ is even. The identity $a^r-1=(a^{r/2}-1)(a^{r/2}+1)$ implies that $N$ divides the product on the right. Minimality of $r$ gives $a^{r/2}\not\equiv1\pmod N$, while the additional condition $a^{r/2}\not\equiv-1\pmod N$ prevents either factor from vanishing modulo $N$. Consequently, at least one of $\gcd(a^{r/2}-1,N)$ and $\gcd(a^{r/2}+1,N)$ is a nontrivial factor of the semiprime. This is the classical conversion used after the order has been recovered.

A good base is a sampled residue $a$ with $\gcd(a,N)=1$, even order $r$, and $a^{r/2}\not\equiv-1\pmod N$. At least one half of the units modulo a product of two distinct odd primes are good~\cite{shor1995factorization}. For a uniformly sampled base, this path contributes at least the probability of sampling a good base multiplied by the conditional probability that the quantum measurement and continued-fraction step recover the relevant order. Independent repetitions then amplify the resulting success probability.

Hadamard gates make the phase register range uniformly over exponents $x$, and controlled modular multiplication computes $a^x\bmod N$ coherently in the work register. Quantum phase estimation and the inverse quantum Fourier transform~\cite{cleve1998revisited,dewolf2019lecturenotes,lin2022qasc,nielsenchuang2010} convert this periodicity into a phase-register measurement near a multiple of $2^t/r$. Continued fractions turn the measured ratio $j/2^t$ into a candidate denominator, and modular exponentiation retains it only if the corresponding power returns to one. Algorithm~\ref{alg:shor-order-finding} gives the complete schedule. The candidate is exact when it equals the least positive period $r$.

\begin{algorithm}[t]
\caption{Quantum order finding with continued-fraction validation~\cite{shor1995factorization}}\label{alg:shor-order-finding}
\SetKwInOut{Input}{Input}
\SetKwInOut{Output}{Output}
\Input{Modulus $N>1$, residue $a$ with $\gcd(a,N)=1$, phase width $t$ satisfying $N^2\leq2^t<2N^2$}
\Output{An accepted positive candidate $r'$, or failure}

Prepare a $t$-qubit phase register in $\ket{0}^{\otimes t}$ and a work register in $\ket{1}$\;
Apply a Hadamard gate to every phase qubit\;
Index the phase qubits by significance, with $i=0$ least significant\;
\For{$i\gets0$ \KwTo $t-1$}{
  Controlled on phase qubit $i$, multiply the work register by $a^{2^i}\bmod N$\;
}
Apply the inverse quantum Fourier transform to the phase register\;
Measure the phase register as $j\in\{0,\ldots,2^t-1\}$\;
Inspect continued-fraction convergents of $j/2^t$ in increasing denominator order\;
\Return{the first positive denominator $r'$ with $a^{r'}\equiv1\pmod N$, or failure}\;
\end{algorithm}

The work register decomposes into eigenstates of modular multiplication, with eigenphases $s/r$ for numerators $s\in\{0,\ldots,r-1\}$. Phase estimation concentrates a measured integer $j$ near $2^t s/r$~\cite{shor1995factorization}. The register window $N^2\leq2^t<2N^2$ and the inequality $r<N$ make recoverable outcomes satisfy $\lvert j/2^t-s/r\rvert<1/(2r^2)$. For $r>0$ and $\gcd(s,r)=1$, this approximation makes $s/r$ a continued-fraction convergent whose reduced denominator is $r$. Exact recovery therefore uses outcomes whose numerators are coprime to $r$; a different sample may produce a multiple of the order or fail the validation check.

There are $\varphi(r)$ numerators in $\{0,\ldots,r-1\}$ that are coprime to $r$, where $\varphi$ is Euler's totient function. Summing their recoverable measurement events and averaging over the eigenphases in the work register yields the lower bound $\varphi(r)/(3r)$ for one order-finding sample. Success means that continued-fraction validation returns the least positive exponent $r$. Reaching a failure probability of at most one half requires repeated samples. The RSA-2048 totals still require a specified number of samples and the corresponding cost within each factoring attempt. The probability bound concerns the measurement distribution, while the inverse quantum Fourier transform contributes separately to the circuit cost.

\subsection{Reversible Modular Arithmetic}

Phase estimation applies modular exponentiation to a superposition of exponents. Temporary values left behind by one exponent would reveal branch information and weaken the interference used to recover the period. When the input registers hold $x$ and $y$ and the scratch space starts at zero, the desired transformation is $\ket{x}\ket{y}\ket{0}\mapsto\ket{x}\ket{a^x y\bmod N}\ket{0}$. Reversible arithmetic achieves this transformation by computing an intermediate value, copying or accumulating the required result, and running the temporary computation backward~\cite{vedral1996arithmetic}.

Here $n$ denotes the word width. Reversible addition retains $u$ and replaces $v$ by $u+v$ in $\mathbb{Z}_{2^n}$; running the same network backward subtracts $u$. A carry flag may be used during the calculation, but it returns to zero before the next operation. The Vedral--Barenco--Ekert network~\cite{vedral1996arithmetic} supplies this basic addition pattern, which later stages reuse for comparison, conditional correction, and uncomputation.

For canonical residues $0\leq u,v<N$, first compute the ordinary sum and compare it with $N$. If the sum reaches $N$, subtract $N$ once; otherwise leave the sum unchanged. Reversing the comparison erases the decision flag while preserving the corrected value, so the target holds $(u+v)\bmod N$ and can immediately feed another arithmetic step. Ref.~\cite{vedral1996arithmetic} gives a reversible network for this sequence.

Let $c$ be a known constant modulo $N$, and let $y_i$ be the bits of $y$. Controlled modular additions accumulate $\left(\sum_i y_i(2^ic\bmod N)\right)\bmod N=cy\bmod N$ in a second register~\cite{vedral1996arithmetic}. When $\gcd(c,N)=1$, the inverse $c^{-1}$ makes this multiplication reversible. After swapping the input and product registers, multiplication by $c^{-1}$ runs backward to erase the old value. The remaining register holds $cy\bmod N$, and the scratch space is ready for the next controlled power.

Algorithm~\ref{alg:controlled-modexp} describes controlled modular exponentiation. Let $\ell$ denote the exponent width, so $x=\sum_{i=0}^{\ell-1}x_i2^i$. The classically precomputed constant $a^{2^i}\bmod N$ is applied exactly when $x_i=1$. The selected constants therefore multiply to $a^x$, yielding $\ket{x}\ket{y}\mapsto\ket{x}\ket{a^x y\bmod N}$. Reversing each multiplier's temporary computation restores the work register.

\begin{algorithm}[t]
\caption{Controlled modular exponentiation for the phase-estimation oracle~\cite{shor1995factorization}}\label{alg:controlled-modexp}
\SetKwInOut{Input}{Input}
\SetKwInOut{Output}{Output}
\Input{Modulus $N$, unit $a\bmod N$, exponent width $\ell$ with bits $x_0,\ldots,x_{\ell-1}$, target $y\in\mathbb{Z}_N$, clean work register}
\Output{Unchanged exponent, target $a^x y\bmod N$, clean work register}

Classically compute $u_i\gets a^{2^i}\bmod N$ for $i=0,\ldots,\ell-1$\;
\For{$i\gets0$ \KwTo $\ell-1$}{
  Controlled on exponent qubit $x_i$, multiply the target by $u_i$ modulo $N$\;
  Uncompute all temporary values produced by the controlled multiplier\;
}
\Return{$x$, $a^x y\bmod N$, and the restored work register}\;
\end{algorithm}

Each controlled multiplication is followed by the inverse operations needed to clear its temporary values. The peak logical-qubit count is set by the data and work registers that must coexist, while Toffoli count and depth accumulate over the controlled multipliers executed in sequence. The powers $a^{2^i}\bmod N$ are computed classically before the quantum circuit begins. The primitive addition and multiplication costs are based on the Vedral--Barenco--Ekert construction~\cite{vedral1996arithmetic}.

\subsection{Eker\aa--H\aa stad Route}

Ref.~\cite{ekera2017quantum} recasts factoring as a short discrete logarithm problem. For this route, let $n$ denote the bit length of $N=pq$, choose a unit $g\in\mathbb{Z}_N^*$ of order $r$, and compute $y=g^{N+1}\bmod N$. Since $N+1-(p+q)=\varphi(N)$ and $r$ divides $\varphi(N)$, the relation $y=g^{p+q}$ holds in the subgroup generated by $g$; when $r>p+q$, the short logarithm is $d=p+q$. The resource presentation in Ref.~\cite{gidney2019factor} uses exponent registers of widths $2m$ and $m$, where $p+q<2^m$ and $m=n/2+O(1)$. The two exponent registers have combined width $3m=1.5n+O(1)$, compared with the $2n$ phase register in the textbook order-finding route. The width is determined from the public bit length of $N$. The hidden factors appear only in the identity that explains why the recovered logarithm equals $p+q$.

The 2017 Eker\aa--H\aa stad formulation~\cite{ekera2017quantum} computes $x=g^{(N-1)/2}$ and recovers the short secret $\delta=(p+q-2)/2$. This parameterization satisfies $d=2\delta+2$ for the sum $d=p+q$ used above. Setting $c=\delta+1$ gives $2c=p+q$, so a candidate root $q'$ is accepted when $q'(2c-q')=N$. The two equations, together with the requirement that the accepted root lie strictly between one and the modulus, make it a proper divisor of $N$. Completing the RSA-2048 analysis requires the probability of obtaining the short secret, the repetitions used to amplify that probability, and the total cost of those repetitions for the Eker\aa--H\aa stad circuit.

Ref.~\cite{gidney2019factor} reports more than $99\%$ classical recovery conditional on a correct short discrete logarithm run. The analysis in Ref.~\cite{shor1995factorization} combines the probability of sampling a good base with repeated order finding to reach high success probability. The success figure used here assumes that each factoring attempt fails with probability at most one half and that three attempts are independent. Their joint failure probability is then at most $1/8\leq1/3$, giving success probability at least $2/3$. Deriving the assumed one-half rate from the one-sample bound above requires an explicit number of order-finding samples and would increase the quantum cost. Ref.~\cite{gidney2019factor} gives $6190$ live logical qubits, $2.7\times10^9$ Toffoli gates, and $2.14\times10^9$ measurement depth for one run. Applying these source figures to three sequential attempts gives $8.1\times10^9$ Toffoli gates and measurement depth $6.42\times10^9$. We separately report a budget of $36906$ abstract classical recovery operations; deriving that budget from Algorithm~\ref{alg:ekera-hastad-rsa} remains open. In the source terminology, the value $2.14\times10^9$ is a measurement depth. Table~\ref{tab:rsa} and the RSA-2048 statement call the tripled value a maximal circuit depth, and relating the two quantities requires an additional argument. This success figure and the reported quantum resources therefore summarize different calculations. A complete analysis must establish both for the Eker\aa--H\aa stad circuit whose resources are reported.

\begin{algorithm}[t]
\caption{Eker\aa--H\aa stad RSA factorization through a short discrete logarithm~\cite{ekera2017quantum,gidney2019factor}}\label{alg:ekera-hastad-rsa}
\SetKwInOut{Input}{Input}
\SetKwInOut{Output}{Output}
\Input{Public RSA modulus $N$, random unit $g\in\mathbb{Z}_N^*$, width $m$ chosen from the public bit length of $N$}
\Output{The factors of $N$, or failure}

Classically compute $y\gets g^{N+1}\bmod N$\;
Prepare uniform exponent registers of widths $2m$ and $m$, together with a clean group register\;
Compute $g^{e_1}y^{-e_2}\bmod N$ in the group register over the exponent superposition\;
Apply the two Fourier transforms and measure a frequency pair\;
Use lattice post-processing to recover $d$ with $0<d<2^m$ and verify $g^d\equiv y\pmod N$\;
Compute $\Delta\gets d^2-4N$ and fail unless $\Delta$ is a nonnegative square\;
\Return{$(d+\sqrt{\Delta})/2$ and $(d-\sqrt{\Delta})/2$}\;
\end{algorithm}

\section{Formalizing P-256}

\subsection{Elliptic-Curve Discrete Logarithms}

Problem~\ref{prob:ecdlp} fixes an elliptic curve $E$ over $\mathbb{F}_p$, a point $P\in E(\mathbb{F}_p)$ of prime order $r$, and a point $Q=[m]P$ for an unknown scalar $m\in\mathbb{Z}_r$. This section works inside the cyclic subgroup $G=\langle P\rangle$. Its computational goal is to recover the unique residue class $m$ represented by the relation $Q=[m]P$.

The underlying discrete logarithm problem is independent of the coordinate representation of the curve. Write the operation of a finite cyclic group additively, let $P$ generate a group $G$ of known order $r$, and write $[a]R$ for the action of $a\in\mathbb{Z}_r$ on $R\in G$. An instance consists of $(G,r,P,Q)$ together with the proposition that some $m\in\mathbb{Z}_r$ satisfies $Q=[m]P$. The success event is that the classical output equals this $m$ in $\mathbb{Z}_r$.

Shor's discrete-logarithm algorithm~\cite{shor1995factorization} uses two exponent registers, a coherent group evaluation, Fourier transforms, and classical post-processing. It is an instance of the abelian hidden-subgroup framework~\cite{childs2021lectures,childsVanDam2010AlgebraicProblems}. For the algebraic derivation, we use the exact order presentation in Ref.~\cite{proos2004shor}; its registers are indexed by $\mathbb{Z}_r$ and its elliptic-curve group action consists of scalar multiplication and point addition. Algorithm~\ref{alg:ecdlp} states this version; later subsections refine its group action step into reversible field and curve circuits.

The coherent map sends $(k,\ell)$ to $[k]P+[\ell]Q=[k+m\ell]P$. Measuring a group element therefore leaves a uniform superposition over one affine fiber of the function $(k,\ell)\mapsto k+m\ell$. The two inverse Fourier transforms map that fiber to frequency pairs satisfying
\begin{equation}
v\equiv mu\pmod r.
\end{equation}
Lean expresses the same relation with the action $[a]P-[b]Q$. Writing its left and right frequencies as $u$ and $w$, respectively, gives $w\equiv-mu\pmod r$. The substitutions $b=-\ell$ and $w=-v$ identify this convention with the additive map in Algorithm~\ref{alg:ecdlp}.

For a valid sample with invertible $u$, the relation gives $m=vu^{-1}$ in $\mathbb{Z}_r$. In the exact prime-order construction presented in Ref.~\cite{proos2004shor}, the first frequency is uniform. Since $r$ is prime, every nonzero frequency is invertible, and these values occupy a fraction $(r-1)/r$ of the frequency space. Our formal proof derives the secret deterministically from a valid linear relation and an invertible frequency. The probability argument is supplied separately: it assumes nonnegative outcome weights and a set of good events whose total weight is at least $2/3$.

To apply the argument for a cyclic group to P-256, we identify the abstract group order, generator, target, and scalar action with their elliptic-curve counterparts. Under these identifications, every certified recoverable measurement outcome gives a candidate for the curve scalar. The current theorem combines three separately provided ingredients: this algebraic identification, correctness and resource bounds for the circuit that evaluates scalar multiples, and the total probability assigned to recoverable measurement outcomes. The formalization appendix maps these ingredients to their Lean declarations. The field and curve circuits needed for the scalar action are developed next.

\begin{algorithm}[t]
\caption{Shor's elliptic-curve discrete-logarithm algorithm~\cite{shor1995factorization,proos2004shor}}\label{alg:ecdlp}
\SetKwInOut{Input}{Input}
\SetKwInOut{Output}{Output}
\Input{Generator $P$ of a cyclic group of known prime order $r$, target point $Q\in\langle P\rangle$}
\Output{The scalar $m\in\mathbb{Z}_r$ satisfying $Q=[m]P$, or failure}

Prepare a uniform superposition over $(k,\ell)\in\mathbb{Z}_r^2$ and a clean group register\;
Coherently compute $[k]P+[\ell]Q$ into the group register\;
Measure and discard the group register\;
Apply the inverse Fourier transform over $\mathbb{Z}_r$ to each index register\;
Measure a frequency pair $(u,v)\in\mathbb{Z}_r^2$\;
\If{$u=0$}{\Return{failure}\;}
\Return{$vu^{-1}\bmod r$}\;
\end{algorithm}

\subsection{Reversible Prime-Field Arithmetic}

Affine elliptic-curve formulas reduce point operations to arithmetic in the prime field $\mathbb{F}_p$. Let $n$ denote the register width; P-256 has $n=256$. We encode a field element by its canonical integer representative in the range $0\leq x<p$. We prove that canonical integers and P-256 field elements encode one another consistently. Each reversible operation preserves its source operands, adds its result to a designated target, and returns flags and scratch registers to their clean values. These properties give point addition a common contract for encoded states.

\begin{algorithm}[t]
\caption{Reversible Montgomery--Kaliski inversion with $2n$ rounds~\cite{roetteler2017ecdlp}}\label{alg:field-inversion}
\SetKwInOut{Input}{Input}
\SetKwInOut{Output}{Output}
\Input{Odd $n$-bit prime $p$, $R=2^n$, nonzero Montgomery representative $\bar a=aR\bmod p$, target representative $\bar z=zR\bmod p$, clean work registers}
\Output{Unchanged $\bar a$, target $\bar z+a^{-1}R\bmod p$, and restored work registers}

Initialize $(u,v,r,s)\gets(p,\bar a,0,1)$, active flag $f\gets1$, padding counter $\ell\gets0$, and clean history bits $m_0,\ldots,m_{2n-1}$\;
\For{$i\gets0$ \KwTo $2n-1$}{
  \eIf{$f=1$ and $v\neq0$}{
    \uIf{$u$ is even}{Set $(u,s)\gets(u/2,2s)$\;}
    \uElseIf{$v$ is even}{Set $(v,r)\gets(v/2,2r)$\;}
    \uElseIf{$u>v$}{Set $(u,r,s)\gets((u-v)/2,r+s,2s)$\;}
    \Else{Set $(v,r,s)\gets((v-u)/2,2r,r+s)$\;}
    Retain the branch selector in $m_i$ and erase the other selector from the parity of $r$ or $s$\;
  }{
    Set $f\gets0$ and $\ell\gets\ell+1$\;
  }
}
Set $k\gets2n-\ell$; then $r\bar a\equiv-2^k\pmod p$\;
Set $\bar b\gets-2^\ell r\bmod p$, so that $\bar b\bar a\equiv R^2\pmod p$ and $\bar b=a^{-1}R\bmod p$\;
Add $\bar b$ to $\bar z$ modulo $p$\;
Reverse the correction and all $2n$ rounds to clean the scratch, counter, flag, and history bits\;
\Return{$\bar a$, $\bar z+a^{-1}R\bmod p$, and the restored work registers}\;
\end{algorithm}

For field elements represented by integers $0\leq a,b<p$, modular addition implements $(a,b)\mapsto(a,a+b)$, with the second component reduced modulo $p$. Ref.~\cite{roetteler2017ecdlp} realizes this map by ordinary addition, conditional subtraction of $p$, and a comparison that erases the reduction flag; reversing the circuit gives subtraction. The corresponding doubling circuit acts on one residue register and uses the oddness of $p$ to recover the reduction decision from the least significant output bit. We prove that the addition and subtraction circuits are mutual inverses on canonical inputs and restore their ancillas. Sequential resource composition takes the maximum workspace and adds Toffoli counts, other gate counts, and depth. Parallel composition adds workspace, Toffoli counts, and other gate counts and takes the maximum depth.

Field multiplication takes two quantum residues and writes their product into a separate target. For the source circuit, set $R=2^n$ and write $\bar a=aR\bmod p$ and $\bar b=bR\bmod p$ for the Montgomery representatives of $a$ and $b$. The construction in Ref.~\cite{roetteler2017ecdlp} maps $\bar a$ and $\bar b$ to $\bar a\bar bR^{-1}=\overline{ab}$, retains one history bit per round recording whether $p$ was added, copies the product, and reverses the computation to erase those bits. The corresponding squaring circuit identifies the two multiplicands, removes the second $n$-qubit input, and adds one copying ancilla, saving $n-1$ qubits. The present P-256 division result uses separately verified circuits for this product, inversion, and target addition. Its correctness and resource bounds depend on all three components.

Ref.~\cite{roetteler2017ecdlp} uses $2n$ Montgomery--Kaliski rounds to map a nonzero representative $\bar a$ to $\overline{a^{-1}}$. The extended binary Euclidean state $(u,v,r,s)$ maintains the B\'ezout relation $p=rv+su$. Each active round retains one selector bit and reconstructs the other from the parity invariant. A flag switches from Euclidean updates to padding rounds, while a counter records their number $\ell$. The final coefficient is an almost inverse. The correction $-2^\ell r\bmod p$ produces a representative $\bar b$ satisfying $\bar b\bar a=R^2\bmod p$, and hence $\bar b=\overline{a^{-1}}$. Algorithm~\ref{alg:field-inversion} adds this value to the target and reverses the recorded computation to restore the workspace.

For a unit denominator $a$ and numerator $b$, modular division adds $ba^{-1}$ to a target. We compose verified circuits for inversion, multiplication, and addition to the target, and then reverse the temporary computations. Under these assumptions, the circuit has a clean quotient action on the P-256 encoding and inherits the component resource bounds. For affine points $(x_1,y_1)$ and $(x_2,y_2)$, generic addition divides by $x_1-x_2$, while point doubling divides by $2y_1$. The mathematical group law handles identity inputs separately. When $x_1=x_2$, equal points use the doubling formula and inverse points sum to the point at infinity; doubling with $y_1=0$ also returns the point at infinity. The generic division result applies after its nonzero denominator has been established.

The current field results therefore cover target addition, inversion, and division under the component assumptions above. They prove clean actions on computational basis inputs and combine the supplied resource bounds. Connecting the concrete Montgomery--Kaliski construction to the assumed P-256 inversion circuit remains a formalization obligation. The formalization appendix gives the corresponding declaration details. The next subsection composes these field operations into elliptic-curve formulas.

\subsection{Reversible Elliptic-Curve Operations}

The mathematical group $E(\mathbb{F}_p)$ contains finite affine points and the point at infinity $\mathcal{O}$. Its addition law treats $\mathcal{O}$ as the identity, sends a pair of inverse points to $\mathcal{O}$, uses a tangent formula when the two inputs coincide, and uses a secant formula when their $x$-coordinates differ. The circuit representation stores only the two coordinates of a finite affine point. Accordingly, the generic affine circuit in Ref.~\cite{roetteler2017ecdlp} takes finite points $P_1=(x_1,y_1)$ and $P_2=(x_2,y_2)$ with $x_1\neq x_2$. Identity, inverse, and doubling cases remain separate branches of the mathematical group law.

For the generic branch, the formulas analyzed in Ref.~\cite{roetteler2017ecdlp} are
\begin{equation}\label{eq:affineaddition}
  \lambda=\frac{y_1-y_2}{x_1-x_2},\qquad
  x_3=\lambda^2-x_1-x_2,\qquad
  y_3=\lambda(x_1-x_3)-y_1.
\end{equation}
They require one field division for the slope, followed by multiplication, squaring, addition, and subtraction. If the curve equation is $y^2=x^3+Ax+B$, nonexceptional doubling replaces the slope by $(3x_1^2+A)/(2y_1)$ and requires $2y_1\neq0$. The Lean mathematical layer proves that both coordinate formulas agree with the corresponding elliptic-curve group operations under their stated denominator conditions. The structured point circuit developed below uses the generic formula in Eq.~\eqref{eq:affineaddition}.

The controlled point addition circuit in Ref.~\cite{roetteler2017ecdlp} keeps $P_1$ in a quantum accumulator and takes $P_2$ as a classically precomputed constant. A control bit selects whether the accumulator becomes $P_1+P_2$ or returns to $P_1$. Reversibility relies on recovering the slope from the output $P_3=(x_3,y_3)$ and the fixed addend:
\begin{equation}\label{eq:recompute-slope}
  \frac{y_1-y_2}{x_1-x_2}=-\frac{y_3+y_2}{x_3-x_2}.
\end{equation}
This identity lets the source circuit update the accumulator directly and erase the slope and coordinate scratch afterward. Our current construction stores the result in separate target coordinates. The construction preserves the input point, computes and writes the generic sum, and reverses the arithmetic to clean the slope and coordinate workspace. Its correctness assumes verified circuits for division and the coordinate updates used in these stages.

For a scalar $k=\sum_{i=0}^{h-1}k_i2^i$, the method for a fixed base in Ref.~\cite{proos2004shor} classically precomputes $P_i=[2^i]P$ and applies one controlled addition of $P_i$ for each bit $k_i$. Thus every elliptic-curve doubling occurs during classical precomputation; the quantum circuit executes only controlled additions of fixed points. Starting from an affine point $S$, the accumulator after the schedule is $S+[k]P$. To verify this schedule, we specify the accumulator before and after every addition controlled by a scalar bit. Every step must satisfy the generic affine condition, including a step whose control bit is zero, and must use a separately verified point addition circuit. Composing these steps proves the final scalar action, workspace cleanup, and accumulated resource bound.

\begin{algorithm}[t]
\caption{Coherent double scalar evaluation~\cite{roetteler2017ecdlp,proos2004shor}}\label{alg:doublescalar}
\SetKwInOut{Input}{Input}
\SetKwInOut{Output}{Output}
\Input{Curve $E/\mathbb{F}_p$ with $n=\lceil\log_2p\rceil$; point $P$ of prime order $r$; point $Q\in\langle P\rangle$; $h$-bit registers $k=(k_i)$ and $\ell=(\ell_i)$ with $h=n+1$; offset $a_0$ sampled uniformly from $\mathbb{Z}_r\setminus\{0\}$; clean point and work registers}
\Output{Unchanged $k,\ell$; point register $[a_0+k]P+[\ell]Q$ when every current accumulator and fixed addend have distinct $x$-coordinates; clean work registers}

Classically precompute $P_i=[2^i]P$ and $Q_i=[2^i]Q$ for $0\leq i<h$\;
Initialize the point register to $[a_0]P$\;
\For{$i\gets0$ \KwTo $h-1$}{
  Coherently add the fixed point $P_i$ under the control bit $k_i$\;
  Erase the slope, coordinate scratch, and arithmetic work registers\;
}
\For{$i\gets0$ \KwTo $h-1$}{
  Coherently add the fixed point $Q_i$ under the control bit $\ell_i$\;
  Erase the slope, coordinate scratch, and arithmetic work registers\;
}
\Return{$k$, $\ell$, $[a_0+k]P+[\ell]Q$, and the clean work registers}\;
\end{algorithm}

The coherent circuit in Ref.~\cite{roetteler2017ecdlp} initializes the accumulator to $[a_0]P$, applies the controlled additions selected by $k$, and then applies those selected by $\ell$ from the precomputed multiples of $Q$. Both exponent registers have width $h=n+1$, and measurement follows all $2h$ controlled additions. Algorithm~\ref{alg:doublescalar} therefore maps each generic basis input to $[a_0+k]P+[\ell]Q$ while preserving coherence across the exponent registers. The fixed offset translates the group value and leaves the hidden period relation detected by the subsequent measurement unchanged~\cite{proos2004shor}. Instantiating two such step lists as a concrete P-256 double scalar circuit and connecting that circuit to the measurement distribution used by scalar recovery remain formalization obligations.

The rough valid-input analysis in Ref.~\cite{roetteler2017ecdlp} chooses $a_0$ uniformly from the nonzero scalars and assumes that the secret scalar is uniform. Under these conditions, the estimated fraction of pairs $(k,\ell)$ for which both schedules avoid exceptional additions is at least $(1-n/2^n)^2$, which becomes $(1-2^{-248})^2$ for P-256. This quantity measures the validity of the generic affine computation. The subsequent measurement and scalar recovery have separate success conditions. Each controlled point addition in the resource analysis uses four inversions, two squarings, and four multiplications~\cite{roetteler2017ecdlp}, with intermediate registers cleaned after each call. Algorithm~\ref{alg:doublescalar} displays the coherent group action with two $h$-bit exponent registers. The $2330$-qubit value in Table~\ref{tab:ecc} comes from the semiclassical construction, which measures and reuses one control qubit after each controlled addition~\cite{roetteler2017ecdlp}. Its Toffoli count and maximal Toffoli-gate depth use the paper's estimate of $2n$ controlled additions. The current resource theorem sums the assumed cost of each controlled point addition in this schedule. A separate probability argument supplies the measurement-outcome weights used for scalar recovery.

\subsection{Scalar Recovery}

The P-256 standard of the National Institute of Standards and Technology~\cite{nist2023sp800186} fixes the prime-field modulus, curve coefficients, base point $P$, prime subgroup order $r$, and cofactor. The current formalization takes as hypotheses the field, curve, and subgroup properties of these constants. The formalization also assumes a correspondence between the generic cyclic group and the P-256 subgroup: their orders agree, their generators and targets correspond to $P$ and $Q$, and their scalar actions agree with elliptic-curve scalar multiplication. Under these hypotheses, a recovered residue gives a candidate solution of $Q=[m]P$. Evidence about the circuit that implements scalar multiplication and its resources is supplied separately from the weights assigned to the full algorithm's measurement outcomes. Deriving both from measurement of a concrete P-256 circuit remains an integration task.

Every certified good frequency pair $(u,v)$ has $u\neq0$ and satisfies the relation $v=mu$ in $\mathbb{Z}_r$ from Algorithm~\ref{alg:ecdlp}. Because $r$ is prime, this nonzero $u$ is invertible. The direct recovery rule in Ref.~\cite{proos2004shor} returns $m'=vu^{-1}$. Multiplying the relation by $u^{-1}$ proves $m'=m$; the correspondence above then gives $Q=[m']P$. The current P-256 theorem begins with a scalar $m$ satisfying $0\leq m<r$ and a proof of $Q=[m]P$. This theorem verifies that every certified good output agrees with the same residue. An execution that receives only the public curve data and computes $m$ remains an integration task.

The probability argument supplies nonnegative outcome weights, a finite set of recoverable outcomes, and a lower bound showing that their total mass is at least $2/3$. One run therefore meets the stated success threshold. The resource analysis in Ref.~\cite{roetteler2017ecdlp} gives the corresponding baseline of $2330$ logical qubits, $1.26\times10^{11}$ Toffoli gates, and maximal Toffoli-gate depth $1.16\times10^{11}$. The seven classical operations in Table~\ref{tab:ecc} form a bookkeeping profile for direct recovery from one certified output: one gcd check, one extended Euclidean run, one modular inversion, one modular addition and one multiplication for checking the relation, and a final negation and multiplication. This count begins after the quantum measurement and excludes candidate search, repeated experiments, Chinese remainder aggregation, and bit complexity. In the current proof route, the measurement weights and resource equalities enter as separate premises.
\section{Conclusion and outlook}

This work develops a Lean formalization of the mathematical reductions and quantum circuit components underlying attacks on RSA-2048 and P-256. For RSA-2048, the formal application layer combines a certified factor, a success threshold of at least $2/3$, and the logical resource values reported in Ref.~\cite{gidney2019factor}. For P-256, it combines a certified private scalar, the same success threshold, and the logical resource values reported in Ref.~\cite{roetteler2017ecdlp}. The two developments bring number theory, probability arguments, reversible arithmetic, quantum circuits, classical recovery, and resource accounting into a common body of checked propositions. Software agents assisted with source analysis, proof decomposition, Lean development, and manuscript maintenance under human review. Machine checking verified every proof accepted into the formal development.

Several connections at the application layer enter as explicit assumptions. The factorization model already includes the two prime factors; the final RSA-2048 proposition then receives factor recovery and success evidence together with resource values imported from the literature. Connecting the formalized Eker\aa--H\aa stad algorithm directly to this proposition remains an integration goal. In the P-256 development, stronger intermediate results use a bridge between the finite cyclic group and the curve, support for the expected measurement outcomes, and facts about the scalar arithmetic circuit. The final P-256 proposition begins with the recovered scalar, its probability bound, and the resource equalities as assumptions. A complete execution beginning from a 2048-bit modulus or a P-256 public point remains to be constructed for each application. The reported resources describe specific analyzed logical circuits; resource lower bounds and optimality remain open questions. For RSA-2048, the source reports logical qubits, Toffoli gates, and measurement depth, while the present proposition uses a general circuit-depth label for that last value. For P-256, the source reports logical qubits, Toffoli gates, and maximal Toffoli-gate depth. The quantitative analysis also counts the stated classical recovery operations. Circuit noise, quantum error correction, magic state production, scheduling, runtime, and physical qubit requirements call for additional models and proofs.

The next stage is to construct executions from public inputs and derive correctness, output probabilities, and resource values from the same formal run. A complementary direction is to extend the resource analysis beyond logical circuits. Recent RSA-2048 estimates study approximate residue arithmetic with lower logical-qubit requirements~\cite{gidney2025less} and fault-tolerant architectures based on quantum low-density parity-check codes~\cite{webster2026pinnacle,cain2026atomic}; formalizing them requires explicit noise, decoding, scheduling, and architecture assumptions. For elliptic-curve discrete logarithms, Ref.~\cite{haner2020improved} develops circuit families optimized separately for width, $T$-gate count, and depth. Formalizing these tradeoffs would extend the single P-256 resource profile used here. Ref.~\cite{luo2026space} develops space-efficient reversible inversion and arithmetic circuits together with a resource derivation. Formalizing its arithmetic circuits and checking the resource derivation would extend the present elliptic-curve analysis. These projects would support comparisons of new quantum cryptanalytic constructions through explicit mathematical assumptions, formal circuit behavior, and mechanically checked resource formulas.

\appendix
\section{Agentic Workflow and Reproducibility}
\label{app:workflow}

\begin{figure}[t]
  \centering
  \includegraphics[width=\linewidth]{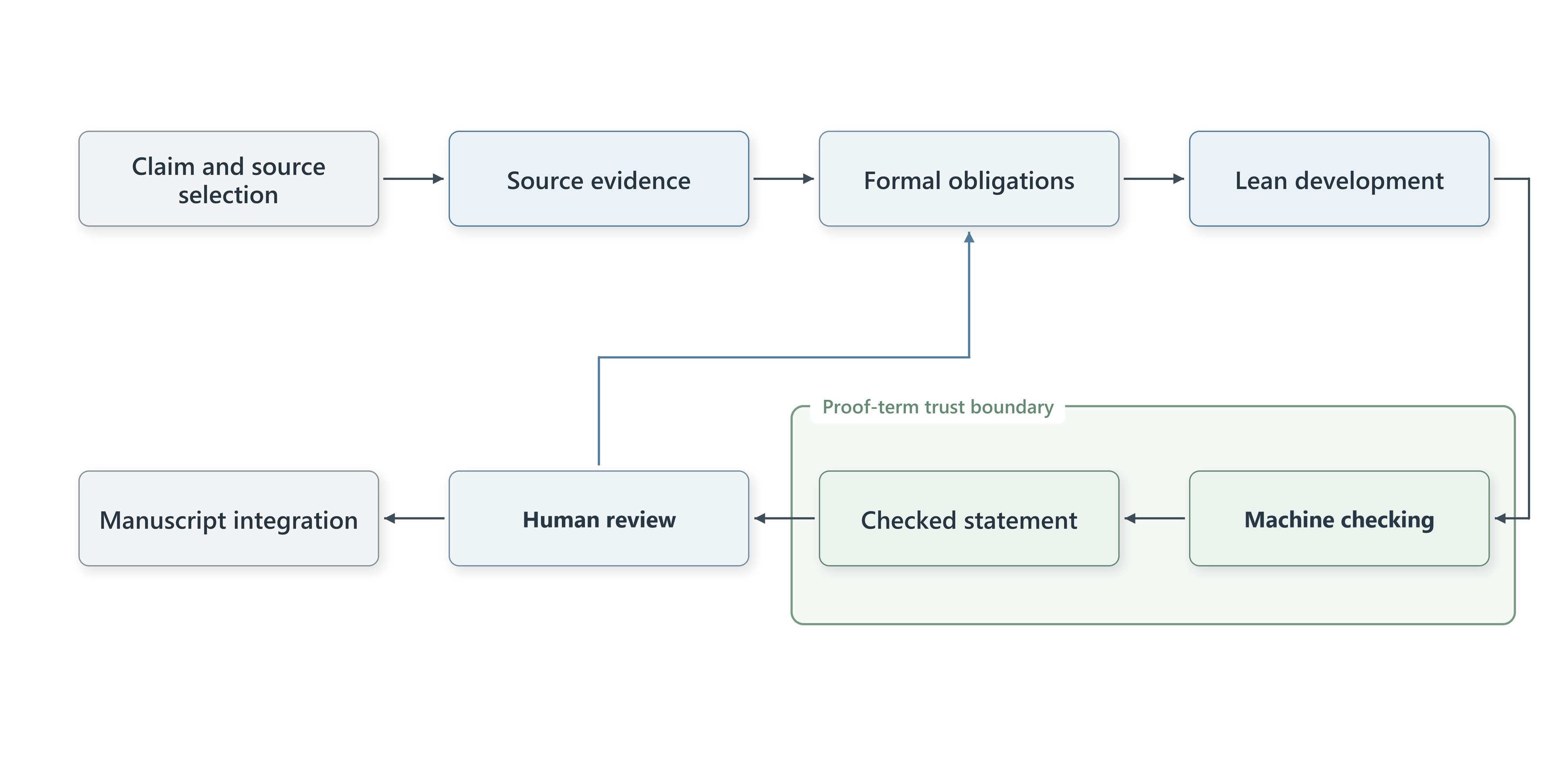}
  \caption{An overview of the agentic formalization workflow and its trust boundary. The two rows follow artifacts from claim selection and source evidence through formal obligations, Lean development, machine checking, human review, and manuscript integration. The framed stages lie within the proof-term trust boundary; the return arrow denotes revision after human review.}
  \label{fig:agentic-workflow}
\end{figure}

\subsection{Responsibilities and Trust Boundary}

Figure~\ref{fig:agentic-workflow} follows each artifact from a scientific question to the manuscript. Authors first specify the claim and the sources that should support it. Software agents retrieve passages, source locations, and metadata, then separate the claim into mathematical, algorithmic, circuit, and resource obligations. They trace relevant dependencies and construct Lean statements and proof terms for those obligations. The kernel accepts a declaration when its proof term checks against the stated type. The accepted theorem type records its explicit hypotheses, while the development preserves imported dependencies for later inspection. Authors receive these checked statements alongside the source evidence and decide which claims, citations, and formal references enter the manuscript. The responsibility of software agents covers evidence retrieval, claim decomposition, dependency tracing, construction of Lean statements and proofs, and an explicit account of the assumptions used at each stage.

Authors retain responsibility for the scientific decisions throughout this flow. They approve the question and source scope, inspect each source passage in context, and decide whether its interpretation supports the proposed mathematical statement. They choose the models and assumptions, compare manuscript prose with the exact checked conclusion, and verify the provenance, units, and scope of quoted resource values. This review also covers citation placement and the prominence given to limitations. A change produced by an agent enters the repository only after an author accepts the corresponding source interpretation, formal statement, and manuscript wording.

The Lean kernel's trust boundary covers the checking of proof terms under the definitions, assumptions, and imported results in the formal environment. Kernel acceptance establishes that a proof term has its declared type. Once accepted, the theorem can be used in later proofs. Source fidelity, model selection, resource inputs drawn from the literature, and the interpretation of a theorem in prose lie in the surrounding research process. Authors review those items by comparing source locations with formal assumptions, checking resource records against their cited analyses, and matching each manuscript claim to the conclusion that the kernel accepted. This division identifies the precise guarantee supplied by formal checking and the evidence that still requires scientific judgment.

\subsection{Positive Collaboration Cases}

For RSA-2048, software agents divided the argument into order finding, reversible modular arithmetic, classical factor recovery, success amplification, and resource accounting for the application claim. They located the source passages and formal declarations for each layer, traced direct dependencies within each branch, and identified the certificates assembled by the terminal theorem. This division let the authors inspect the number-theoretic reduction separately from the circuit construction. The authors also reviewed the success certificate and the literature resource tuple as separate inputs, checking the provenance and scope of each before accepting the application statement. The declaration roadmap records where every component enters the endpoint.

For P-256, software agents mapped the generic recovery argument for a finite cyclic group alongside the reversible prime-field and elliptic-curve circuit results. They recorded the equations and certificates required to assemble the application endpoint, including the curve bridge, sampling support, certificate for the scalar oracle, and resource equalities. Authors checked the endpoint's output equality against the stored private scalar and separately checked the provenance and scope of the logical resource tuple. The terminal theorem receives these correctness and resource facts as explicit inputs. Appendix~\ref{app:lean-formalization} gives a reviewable account of their mathematical roles and current connections.

\subsection{Reproducibility}

A reader can start from the two theorem environments in Sec.~\ref{sec:main-results} and follow Appendix~\ref{app:lean-formalization} to the declaration roadmap and terminal signatures in Listings~\ref{lst:rsa-terminal} and~\ref{lst:p256-terminal}. The exact source versions and locators are recorded in \nolinkurl{registry/sources.jsonl}. For RSA-2048, Shor's construction~\cite{shor1995factorization} supports the factoring route, while Ref.~\cite{gidney2019factor} supplies the quoted quantum resource tuple. The success certificate and classical operation count are separate repository inputs. These inputs, the assumptions and conclusion of the final Lean theorem, the RSA-2048 theorem, and Table~\ref{tab:rsa} must agree. For P-256, National Institute of Standards and Technology Special Publication 800-186~\cite{nist2023sp800186} supplies the domain parameters, while Ref.~\cite{roetteler2017ecdlp} supplies the quoted quantum resource tuple. The success certificate and the count of seven classical operations are separate repository inputs. These inputs must agree with the assumptions and conclusion of the final Lean theorem, the P-256 theorem, and Table~\ref{tab:ecc}. Agreement among the source records, checked propositions, and manuscript claims is the reproducibility criterion for each application result.
\section{Lean Formalization}
\label{app:lean-formalization}

\subsection{Formalization Roadmap}

\begin{figure}[t]
  \centering
  \includegraphics[width=\linewidth]{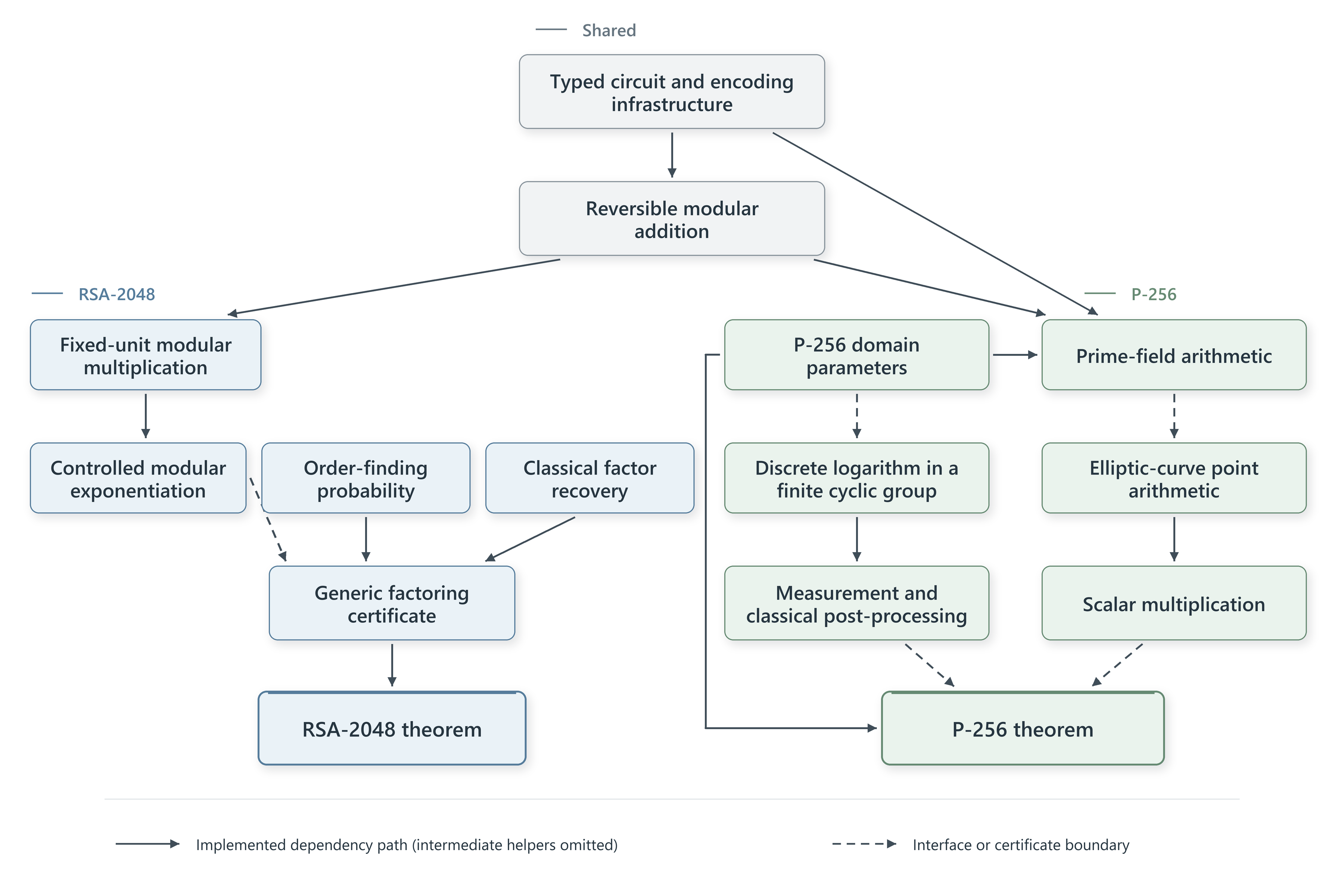}
  \caption{A dependency map of the Lean development used for the RSA-2048 and P-256 results. Solid arrows show implemented dependency paths, with intermediate helpers omitted; dashed arrows show interface or certificate boundaries. The terminal nodes are the public RSA-2048 and P-256 theorems.}
  \label{fig:formalization-dag}
\end{figure}

This appendix maps the mathematical development to selected Lean declarations. Figure~\ref{fig:formalization-dag} separates implemented dependency paths from interfaces that accept an explicit bridge or certificate. The branch for a 2048-bit Rivest--Shamir--Adleman (RSA) modulus joins reversible arithmetic with probability and factor recovery at a generic factoring certificate. The P-256 branch joins discrete logarithms in finite cyclic groups, field and curve arithmetic, and scalar multiplication at a witness in the application layer. The two terminal propositions record the returned value, a rational success bound, and logical resource values derived from the literature. A solid arrow means that a declaration directly constructs or consumes the preceding object. A dashed arrow means that the receiving theorem asks its caller to provide the stated relation as a hypothesis or certificate field.

Typed circuits carry an encoding, a state transformation, and resource fields through composition. Modular addition is a representative foundation: on a clean basis state it preserves the left register, replaces the right register by its sum modulo the declared modulus, and returns the flag to zero. Listing~\ref{lst:modular-addition} gives this semantic statement; the proof body and the accompanying resource witness are available in the repository.

\begin{leanlisting}{The clean basis state semantics of reversible modular addition. The proof body is omitted.}{lst:modular-addition}
theorem addCircuit_apply_clean_ket (N n : ℕ) [NeZero N]
    (params : ResourceParameters) (a b : ZMod N) :
    Circuit.apply (addCircuit N n params)
      (PureState.ket (R := register N)
        ({ left := a, right := b, flag := false } : Data N) :
          StateVector (register N)) =
      (PureState.ket (R := register N)
        ({ left := a, right := a + b, flag := false } : Data N) :
          StateVector (register N)) := by
  -- proof omitted
\end{leanlisting}

The listing also illustrates how local semantic and resource claims are organized. Its state equality describes the behavior visible to a caller, including cleanup of the ancillary flag. A companion witness refers to this same circuit when it projects qubits, gates, depth, and query depth. Composition preserves this association as addition feeds multiplication and exponentiation. At the ends of the graph, a broader proposition collects such circuit results with probability and number-theoretic certificates. The declaration tables below give a compact route from each part of this narrative to the relevant namespace.

\subsection{RSA-2048 Declarations}

The factorization development starts from a semiprime model that stores two distinct prime factors, their product, and the consequences needed to recognize a nontrivial divisor. Order finding uses the multiplicative order supplied by Mathlib~\cite{mathlib}. For a sampled unit, the Chinese remainder decomposition relates this order to the orders modulo the two prime factors. Their powers of two determine the event needed by classical factor recovery. Under the stated premise on the prime factors, a counting theorem shows that at least half of the units satisfy that condition. The probability layer assigns an exact weight to every pair of phase and orbit outcomes. It then restricts the sum to measurements whose continued-fraction candidate equals the true order and proves a rational lower bound for this event. The generic factoring certificate receives the resulting success data and selects a returned divisor from the pair carried by the semiprime model.

The arithmetic path composes reversible addition, multiplication by a fixed unit, and controlled modular exponentiation. The Vedral--Barenco--Ekert construction~\cite{vedral1996arithmetic} supplies the reference progression from addition to modular exponentiation. In Lean, multiplication by a fixed unit first accumulates a scaled input, swaps the two data registers, and subtracts a scaled value using the inverse unit. On a clean input, this sequence leaves the product in the multiplicand register and clears the accumulator and flag. A schedule for modular exponentiation applies controlled powers of the fixed base, with an inverse step available for cancellation and cleanup. Correctness, inverse behavior, control branches, and resource projections refer to the selected circuit objects. Listing~\ref{lst:modular-multiplication} records the semantics on a clean input that underlie the certified multiplication circuit. The complete circuit action and its resource witness are separate declarations.

\begin{leanlisting}{The semantics on a clean input that underlie the certified multiplication circuit. The definition and theorem statement are shown, and the proof is omitted. The complete \texttt{Circuit.apply} theorem lies outside this excerpt.}{lst:modular-multiplication}
def macSwapUncompute {N : ℕ} (u : (ZMod N)ˣ)
    (x : ModularMultiplyAccumulate.Data N) :
    ModularMultiplyAccumulate.Data N :=
  (swapMacRegisters (x.addScaled (u : ZMod N))).subScaled
    ((u⁻¹ : (ZMod N)ˣ) : ZMod N)

theorem macSwapUncompute_cleanInput_multiplyByUnit {N : ℕ}
    (u : (ZMod N)ˣ) (x : ZMod N) :
    macSwapUncompute u (cleanInput x) =
      ({ multiplicand := multiplyByUnit u x,
         accumulator := 0,
         flag := false } : ModularMultiplyAccumulate.Data N) := by
  -- proof omitted
\end{leanlisting}

The Eker\aa--H\aa stad route~\cite{ekera2017quantum} has its own short discrete logarithm certificate. The certificate accepts the mass of the recovery event and a bounded classical search witness. The classical layer recovers a half sum, checks the associated quadratic relation, and proves that the returned value belongs to the stored factor pair. Its packaged theorem exposes correctness, success, repetition, and resource statements for this route. The public RSA-2048 endpoint selects the Shor route factoring certificate, fixes three trials, and attaches a logical resource record. Evaluating and rounding the formulas and table in Ref.~\cite{gidney2019factor} gives the values for one run: $6190$ logical qubits, $2.7\times10^9$ Toffoli gates, and $2.14\times10^9$ measurement depth.

The public RSA proposition expands these inputs in Listing~\ref{lst:rsa-terminal}. Its input includes the semiprime factor model and a proof that the modulus lies in the declared 2048-bit interval. The existential output equals one of the stored factors. One rational inequality records the lower bound of $2/3$ after three trials, and a second relates that probability to the final failure budget. The retry certificate sets the run count to three. The resulting totals are $8.1\times10^9$ Toffoli gates and $6.42\times10^9$ units of the source measurement depth~\cite{gidney2019factor}; Lean stores the latter value in the general \texttt{circuitDepth} field. The proposition also records $36{,}906$ classical arithmetic operations and two readiness flags supporting the values summarized in Table~\ref{tab:rsa}.

\begin{leanlisting}{The RSA-2048 public proposition. It records factor recovery, rational success bounds, three-run accounting, and logical resources.}{lst:rsa-terminal}
def PublicTheoremShape (input : InputClass) : Prop :=
  ∃ d successNumerator successDenominator : ℕ,
    0 < successDenominator ∧
      (d = input.model.leftFactor ∨
       d = input.model.rightFactor) ∧
      2 * successDenominator <= 3 * successNumerator ∧
      (failureBudget.failureDenominator -
          failureBudget.failureNumerator) * successDenominator <=
        failureBudget.failureDenominator * successNumerator ∧
      2 ^ (modulusBits - 1) <= input.modulus ∧
      input.modulus < 2 ^ modulusBits ∧
      retry.runCount = 3 ∧
      finalSuccessAccountedEnvelope.fields.logicalQubits = 6190 ∧
      finalSuccessAccountedEnvelope.fields.toffoliGates =
        retry.runCount * 2700000000 ∧
      finalSuccessAccountedEnvelope.fields.circuitDepth =
        retry.runCount * 2140000000 ∧
      classicalArithmeticOps = 36906 ∧
      retry.readyForFinalStatement = true ∧
      finalSuccessAccountedEnvelope.readyForFinalStatement = true
\end{leanlisting}

\begin{table}[t]
  \centering
  \caption{A map from selected Lean declarations to the RSA construction in the manuscript. Here declarations are grouped by namespace, and each entry in the first column omits the displayed namespace prefix.}
  \small
  \setlength{\tabcolsep}{0.6em}
  \begin{tabularx}{\linewidth}{>{\raggedright\arraybackslash}p{0.3\linewidth}>{\raggedright\arraybackslash}X}
    \toprule
    Declaration & Role in the manuscript \\
    \midrule
    \multicolumn{2}{@{}p{\linewidth}@{}}{\nolinkurl{OrderFinding.ShorSourceJoint}} \\
    \nolinkurl{main} & Bounds the probability of recovering the true order from validated outcomes. \\
    \midrule
    \multicolumn{2}{@{}p{\linewidth}@{}}{\nolinkurl{ShorFactoring.SemiprimeFactorModel}} \\
    \nolinkurl{crtGoodBaseCount_atLeastHalf} & Counts bases that satisfy the factor recovery condition. \\
    \midrule
    \multicolumn{2}{@{}p{\linewidth}@{}}{\nolinkurl{ModularExponentiation}} \\
    \nolinkurl{controlledPowerScheduleCircuit} & Builds the controlled powers used by order finding. \\
    \midrule
    \multicolumn{2}{@{}p{\linewidth}@{}}{\nolinkurl{Factoring.EkeraHastadStyle}} \\
    \nolinkurl{main_factorization} & Packages the short discrete logarithm factorization route. \\
    \midrule
    \multicolumn{2}{@{}p{\linewidth}@{}}{\nolinkurl{Factoring.FormulaEnvelope.RSA2048}} \\
    \nolinkurl{retryAmplificationCertificate} & Fixes the success accounting for three runs. \\
    \nolinkurl{finalLogicalResourceEstimate} & Keeps the logical qubit count fixed and scales gates and depth to three runs. \\
    \nolinkurl{main} & Proves the public RSA-2048 proposition. \\
    \bottomrule
  \end{tabularx}
\end{table}

\subsection{P-256 Declarations}

The P-256 development fixes the field modulus, curve coefficients, base point, subgroup order, and cofactor specified by the National Institute of Standards and Technology~\cite{nist2023sp800186}. Field elements are residues modulo the standardized prime, and finite points carry coordinates together with the curve equation. The application carrier supplies a public key, a private scalar with a range proof, and an abstract scalar multiplication map that relates them. A bridge interprets this curve data as a discrete logarithm problem in a finite cyclic group. It equates the relevant orders, maps the generator and target to the base point and public key, and preserves scalar multiplication. Probability bounds are rational inequalities. The final resource record contains logical qubits, Toffoli gates, maximal Toffoli depth, and classical operations.

Shor's discrete logarithm construction~\cite{shor1995factorization} uses two exponent registers. For exponents $a$ and $b$, its group action multiplies the workspace by the generator raised to $a$ and the target raised to $-b$. Fourier readout of both exponent registers produces a pair of frequencies. A source relation characterizes valid pairs by a linear congruence, and the classical recovery step solves that congruence when the selected frequency is invertible. Ref.~\cite{roetteler2017ecdlp} realizes this architecture with controlled elliptic-curve additions. In Lean, a support certificate supplies the weight of each outcome, the selected recovery event, proofs that its outcomes satisfy the source relation, and a rational lower bound on its total weight. The Fourier readout circuit and these outcome weights enter through supplied fields. The bridge from the cyclic group to the curve and the scalar circuit are further inputs to the recovery theorem.

The arithmetic branch begins with witnesses for field addition, inversion, and division on clean inputs. Each witness states the encoded result, cleanup of its work registers, and the resources of the supplied program. Division also receives a multiplication program and a proof that it realizes the required field product. Affine point addition then consumes coordinate programs under the nonexceptional condition for the selected formula. The controlled version proves the identity action when its control is zero and the prescribed point addition when the control is one. Scalar multiplication folds these controlled steps over a bit schedule. Its run object carries the accumulator trace and the validity premise for every scheduled addition. The P-256 field bundle and the point arithmetic witness occupy separate interfaces. The structured scalar endpoint associates its action on basis states and resource profile with one folded circuit.

\begin{table}[t]
  \centering
  \caption{A map from selected Lean declarations to the P-256 construction in the manuscript. Here declarations are grouped by namespace, and each entry in the first column omits the displayed namespace prefix.}
  \small
  \setlength{\tabcolsep}{0.6em}
  \begin{tabularx}{\linewidth}{>{\raggedright\arraybackslash}p{0.3\linewidth}>{\raggedright\arraybackslash}X}
    \toprule
    Declaration & Role in the manuscript \\
    \midrule
    \multicolumn{2}{@{}p{\linewidth}@{}}{\nolinkurl{P256PrimeFieldBridge.WitnessBundle}} \\
    \nolinkurl{inversionResourceCorrectWitness} & Projects inversion semantics on clean inputs and resources. \\
    \nolinkurl{divisionResourceCorrectWitness} & Projects division semantics on clean inputs and resources. \\
    \midrule
    \multicolumn{2}{@{}p{\linewidth}@{}}{\nolinkurl{PrimeFieldShortWeierstrass.ControlledPointAddition}} \\
    \nolinkurl{DecomposedWitness} & Certifies controlled affine point addition. \\
    \midrule
    \multicolumn{2}{@{}p{\linewidth}@{}}{\nolinkurl{PrimeFieldShortWeierstrass.ScalarMultiplication.}\newline\nolinkurl{CertifiedEndpoint.ComposedEndpoint}} \\
    \nolinkurl{Witness} & Associates a scalar schedule with one circuit and its resources. \\
    \midrule
    \multicolumn{2}{@{}p{\linewidth}@{}}{\nolinkurl{P256LogicalResources.SameCircuitScalarRecovery}} \\
    \nolinkurl{Witness} & Collects the supplied bridge, probability, circuit, and resource certificates. \\
    \midrule
    \multicolumn{2}{@{}p{\linewidth}@{}}{\nolinkurl{P256LogicalResources.PublicTheoremShape}} \\
    \nolinkurl{main} & Constructs the public P-256 proposition. \\
    \bottomrule
  \end{tabularx}
\end{table}

At the application layer, the comprehensive recovery witness accepts the curve bridge, Fourier support, scalar oracle certificate, output equalities, and public resources as supplied fields. The scalar oracle certificate links its clean basis action and resources to one circuit object. The Fourier support supplies the probability data used to choose and validate the classical output. The public carrier supplies the public point, the private scalar, and the abstract scalar multiplication map; the endpoint identifies the returned natural number with that carried scalar. A candidate proof checks the same value against the induced curve problem, and a range proof places it below the subgroup order. Listing~\ref{lst:p256-terminal} expands this proposition. Its success inequality states a lower bound of $2/3$, and both run multipliers equal one. The resource tuple is $2330$ logical qubits, $1.26\times10^{11}$ Toffoli gates, and $1.16\times10^{11}$ maximal Toffoli depth from Ref.~\cite{roetteler2017ecdlp}. Seven classical operations come from the local recovery accounting. These fields correspond to Table~\ref{tab:ecc}, and the readiness flag marks the completed numerical record consumed by the public proposition.

\begin{leanlisting}{The P-256 endpoint statement and its existential public proposition. They record scalar recovery, a rational success bound, and logical resources.}{lst:p256-terminal}
namespace PublicEndpointWitness

def Statement (I : P256DomainParameters.ScalarRecoveryInterface)
    (w : PublicEndpointWitness I) : Prop :=
  w.output = I.privateScalar.value ∧
    w.output < P256DomainParameters.subgroupOrder ∧
    I.toECDLPInstance.CandidateScalar
      (w.output : ZMod I.toECDLPInstance.subgroupOrder) ∧
    2 * w.successDenominator <= 3 * w.successNumerator ∧
    0 < w.successDenominator ∧
    w.resources.logicalQubits = 2330 ∧
    w.resources.toffoliGates = 126000000000 ∧
    w.resources.maximalToffoliDepth = 116000000000 ∧
    w.resources.classicalOps = 7 ∧
    w.resources.toffoliRunMultiplier = 1 ∧
    w.resources.depthRunMultiplier = 1 ∧
    w.resources.readyForFinalStatement = true

end PublicEndpointWitness

def PublicTheoremShape
    (I : P256DomainParameters.ScalarRecoveryInterface) : Prop :=
  ∃ witness, PublicEndpointWitness.Statement I witness
\end{leanlisting}

\subsection{Main Result Assembly}

The two branches use the same assembly pattern at their lower circuit layers: a correctness theorem and a resource witness refer to one circuit object. Their public propositions operate at a higher layer. The RSA wrapper receives a semiprime model and internally selects the factor and success certificate, retry accounting, and numerical resource fields. Its values for three runs scale an estimate for one run derived from Ref.~\cite{gidney2019factor}. The factor already belongs to the semiprime model, and the probability is represented by exact ratios of natural numbers. The P-256 wrapper receives a carrier that includes the private scalar, together with a candidate proof, a rational success bound, and resource equalities. The comprehensive recovery witness can collect the curve bridge, outcome support, scalar circuit, and resource record before this final step. Each endpoint states exactly which evidence the application theorem consumes.

The terminal statements certify conjunctions over mathematical and numerical evidence assembled in the application layer. The circuit declarations prove local actions on basis states, cleanup, control behavior, and resources. Number-theoretic certificates turn valid quantum outcomes into factors or scalars. Bridge records connect finite cyclic groups to elliptic-curve scalar multiplication, and support records provide the probability data used by classical recovery. This separation makes the input to each theorem visible in its type and preserves the local association between circuit behavior and circuit resources.

Readers can audit the construction at three levels. First, the arithmetic statements expose the action of each circuit on a clean encoded state and the cleanup of work registers. Second, the probability and recovery statements name the selected event and the algebraic condition that turns an outcome into a factor or scalar. Third, the terminal propositions display every equality used to report success and resources. On the RSA circuit branch, this path runs from modular addition through exponentiation to the factoring certificate. On the P-256 branch, it runs from field programs through controlled point addition and the scalar schedule to the supplied recovery witness. The declaration tables locate both endpoints. Listings~\ref{lst:rsa-terminal} and~\ref{lst:p256-terminal} make explicit the certificate boundary behind the numerical claims reported in Sec.~\ref{sec:main-results}.

\end{document}